\documentclass[11pt,a4paper,onecolumn,reqno]{amsart}
\usepackage[a4paper, margin=2.3cm, bmargin=3cm]{geometry}

\usepackage{graphicx,stfloats} \usepackage{color}
\usepackage{amsmath}
\usepackage{amsaddr}
\usepackage{bm, soul}
\usepackage{mathtools}
\usepackage[colorinlistoftodos,prependcaption,textsize=tiny]{todonotes}
\usepackage{braket}
\usepackage{hyperref}
\usepackage[square,comma,sort&compress, numbers]{natbib}
\usepackage{enumerate}
\usepackage{enumitem}
\usepackage{floatrow}
\usepackage{tikz}
\usepackage{pgfplots}
\usepackage{csvsimple}
\usepackage{pict2e}
\usepackage{overpic}
\pgfplotsset{compat=1.9}
\usetikzlibrary{calc,arrows,arrows.meta,fadings,decorations.pathreplacing,decorations.markings,patterns,shapes.geometric}
\tikzset{>=stealth,inner sep=0pt, outer sep=2pt,}
\tikzset{vecteur/.style={->,thick,color=black,smooth}}

\usepackage{placeins}

\renewcommand{\st}[1]{}

\usepackage[version=3]{mhchem} \usepackage{siunitx}
 \usepackage{multirow}

\usepackage{etoolbox}
\patchcmd{\pprintMaketitle}
  {\hrule\vskip12pt}
 {\hrule\vskip12pt\ifvoid\extrainfobox\else\unvbox\extrainfobox\par\vskip12pt\fi}
 {}{}

\newsavebox\extrainfobox

\usepackage{caption}
\title{
    Flame-wall interaction of thermodiffusively unstable hydrogen/air flames -- Part I: Characterization of governing physical phenomena
}

\author[stfs]{Max Schneider$^{*}$, Hendrik Nicolai, Vinzenz Schuh, Matthias Steinhausen, Christian Hasse}
\email{schneider@stfs.tu-darmstadt.de} 
\address[]{Technical University of Darmstadt, Department of Mechanical Engineering, Simulation of reactive Thermo-Fluid Systems, Otto-Berndt-Stra{\ss}e 2, 64287 Darmstadt, Germany\\
}

\begin{document}
\pagestyle{plain}

\maketitle

\begin{abstract} 
Hydrogen combustion systems operated under fuel-lean conditions offer great potential for low emissions.
However, these operating conditions are also susceptible to intrinsic thermodiffusive combustion instabilities. 
Even though technical combustors are enclosed by walls that significantly influence the combustion process, intrinsic flame instabilities have mostly been investigated in canonical freely-propagating flame configurations unconfined by walls.
This study aims to close this gap by investigating the flame-wall interaction of thermodiffusive unstable hydrogen/air flame through detailed numerical simulations in a two-dimensional head-on quenching configuration. 
It presents an in-depth qualitative and quantitative analysis of the quenching process, revealing the major impact factors of the instabilities on the quenching characteristics. 
The thermodiffusive instabilities result in lower quenching distances and increased wall heat fluxes compared to one-dimensional head-on quenching flames under similar operation conditions. 
The change in quenching characteristics seems not to be driven by kinematic effects.
Instead, the increased wall heat fluxes are caused by the enhanced flame reactivity of the unstable flame approaching the wall, which results from mixture variations associated with the instabilities.
Overall, the study highlights the importance of studying flame-wall interaction in more complex domains than simple one-dimensional configurations, where such instabilities are inherently suppressed. 
Further, it emphasizes the need to incorporate local mixture variations induced by intrinsic combustion instabilities in combustion models for flame-wall interactions.
In part II of this study, the scope is expanded to gas turbine and internal combustion engine relevant conditions through a parametric study, varying the equivalence ratio, pressure, and unburnt temperature.
\end{abstract}

\keywords{\textbf{Keywords:} Flame-wall interaction; Thermodiffusive instability; Head-on quenching; Hydrogen; Premixed }

\section*{Novelty and Significance Statement}
This work presents novel simulations and in-depth analysis of flame-wall interaction (head-on quenching) of laminar thermodiffusively unstable hydrogen/air flames.
Thermodiffusive instabilities are significant in technical combustion chambers enclosed by walls, such as gas turbines and internal combustion engines, particularly under lean conditions, where they raise safety concerns like flame flashback and increased thermal loads on walls.
The study shows that these instabilities strongly affect flame-wall interactions, leading to smaller quenching distances and higher wall heat fluxes than in one-dimensional head-on quenching.
Consequently, this work demonstrates that for flames susceptible to instabilities, such as lean hydrogen/air flames, one-dimensional head-on quenching simulations are inadequate for accurately determining wall heat fluxes and quenching distances.
Additionally, this study highlights that differential diffusion effects induced by the intrinsic instabilities must be considered in combustion models, requiring full resolution of the local mixture distribution.

\section*{CRediT authorship contribution statement}
\textbf{Max Schneider}: Conceptualization, Methodology, Investigation, Software, Formal analysis, Data Curation, Visualization, Writing - original draft.
\textbf{Hendrik Nicolai}: Conceptualization, Methodology, Supervision, Project administration, Writing - review \& editing.
\textbf{Vinzenz Schuh}: Software, Formal analysis, Writing - review \& editing.
\textbf{Matthias Steinhausen}: Conceptualization, Methodology, Supervision, Writing - review \& editing.
\textbf{Christian Hasse}: Resources, Supervision, Project administration, Funding acquisition, Writing - review \& editing.

\section{Introduction} \addvspace{10pt}
\label{sec:introduction}
\setlist{nolistsep}
For the transition to a carbon-free energy system, chemical energy carriers such as green hydrogen (\ce{H2}) are essential~\cite{Dreizler2021}.
In particular, hydrogen has significant potential to decarbonize current technologies, for example in the transport sector, industrial processes, and power supply.
One potential usage of hydrogen is as a fuel~\cite{Verhelst2009}.
In this context, the most promising approaches for hydrogen combustion involve fuel-lean premixed combustion systems.
These systems produce lower \ce{NOx} emissions compared to their non-premixed counterparts due to the lower exhaust gas temperatures~\cite{Correa1993}.
Besides the advantages, fuel-lean hydrogen combustion presents additional challenges in the design of novel combustion systems.
One of these challenges is the susceptibility of lean-premixed hydrogen/air flames to intrinsic combustion instabilities, which can significantly impact the combustion characteristics.
These intrinsic instabilities result from the superposition of two distinct effects:
\begin{itemize}[noitemsep]
    \item \textbf{Thermodiffusive instability}:
    Hydrogen's low Lewis number, resulting from its high molecular diffusivity relative to the thermal diffusivity of the mixture, leads to strong differential diffusion effects.
    The resulting imbalance between heat and hydrogen mass fluxes amplifies small spatial perturbations in the flame front, causing strongly wrinkled flame fronts and significantly increasing fuel consumption rates.
    \item \textbf{Hydrodynamic instability}:
    Additionally, the flame propagation is influenced by hydrodynamic instability originating from the density jump across the flame front which is a destabilizing effect in all premixed flames, regardless of the mixture's Lewis number.
\end{itemize}
Understanding the impact of the instabilities on flame dynamics, heat release rates, and flame consumption speeds is critical for safety and thermal efficiency.
Therefore, intrinsic instabilities were recently investigated through experimental investigations \cite{Kwon2002, Bauwens2017, FernandezGalisteo2018}, asymptotic analysis \cite{Creta2020}, and direct numerical simulations (DNS) \cite{Altantzis2011, Altantzis2012, Frouzakis2015, Berger2019, Attili2021, Howarth2022, Wen2022a, Wen2022b, Boettler2024, Schuh2024}. 
For a more comprehensive overview of the current research endeavours on thermodiffusive instabilities in lean hydrogen/air flames, the reader is referred to the review paper by Pitsch \cite{Pitsch2024}.
Collectively, the studies demonstrate that the increase in the overall consumption speed associated with the combustion instabilities can be attributed to a combination of an enlarged flame surface area as well as an increased fuel consumption rate per unit of flame surface area \cite{Altantzis2015, Frouzakis2015, Berger2019, Creta2020, Attili2021, Howarth2022}.
Particularly relevant to this study, a common approach in the numerical investigations is the modification of domain dimensions to alter the wavelengths and cell size distributions of the instabilities.
This approach is applied in both the investigation of the early (linear regime) and long-term (non-linear regime) evolution of flame dynamics in unstable flames:
\begin{itemize}[noitemsep]
    \item Regarding early flame propagation, the influence of hydrogen’s low Lewis number on the linear stability of the planar flame can be characterized by a dispersion relation that describes the growth rate of small perturbations as a function of the wavenumber of the perturbation.
    This dispersion relation can be estimated using theoretical models \cite{Matalon2003} or numerically evaluated through simulations \cite{Berger2019, Lapenna2019, Lulic2023, Pitsch2024} in domains of varying heights, modifying the cell size distributions of the instabilities.
    \item Regarding the long-term non-linear evolution of the flame dynamics, Altantzis et al.~\cite{Altantzis2012} and Frouzakis et al.~\cite{Frouzakis2015} studied premixed hydrogen/air flames in two-dimensional channel-like domains.
    They found that the flame dynamics depend strongly on the cell size distribution and the Lewis number.
    For instance, they observed that the propagation speed increases with increasing lateral domain size and asymptotically approaches values larger than the laminar flame speed.
    Building on these findings, Berger et al.~\cite{Berger2019} showed that flame dynamics become independent of domain size in sufficiently large domains.
    However, in domains narrower than approximately $100$ thermal flame thicknesses, flame consumption speed, and nonlinear dynamics are significantly affected by the domain size \cite{Frouzakis2015, Altantzis2012}.
    This is attributed to the influence of domain height on the relative importance of thermodiffusive and hydrodynamic instabilities \cite{Altantzis2012, Schuh2024}.
\end{itemize}

All of the aforementioned studies on thermodiffusive instabilities focus on freely-propagating flames in open domains without considering the presence of combustor walls.
However, Welch et al.~\cite{Welch2024} recently observed thermodiffusive combustion instabilities in their experimental study of a spark-ignition hydrogen engine operating under lean conditions.
These findings indicate that thermodiffusive combustion instabilities are relevant for practical combustion systems enclosed by walls.
In these systems, the interaction of the flame with the combustor wall introduces an additional level of complexity \cite{Dreizler2015}.
One of the governing phenomena are flame-wall interactions (FWIs) that lower the overall combustion efficiency~\cite{Dreizler2015}, impact pollutant formation \cite{Lai2018, Steinhausen2023} and can also lead to undesired flame behavior, such as flashback \cite{Fritz2004}.
FWI is especially relevant for hydrogen flames, since its high reactivity causes it to burn closer to walls and, therefore, leads to higher wall heat fluxes and thermal loads compared to hydrocarbons \cite{Dabireau2003, Gruber2010, DeNardi2024}.

To gain a deeper understanding of the governing phenomena, FWI is typically studied in Head-On Quenching (HOQ) - a transient process where a flame front approaches a wall and extinguishes over time - and Side-Wall Quenching (SWQ) - where the flame quenches in the boundary layer flow, leading to a quasi-steady state - configurations, both in laminar and turbulent flows~\cite{Dreizler2015}.
The majority of FWI studies focused on methane and higher hydrocarbons; a detailed overview can be found in the review paper by Dreizler and Böhm \cite{Dreizler2015}.
Concerning hydrogen flames, Dabireau et al.~\cite{Dabireau2003} and Gruber et al.~\cite{Gruber2010} both examined FWI of hydrogen/air flames, however, under conditions that are not susceptible to thermodiffusive instabilities.
Recently, there has been a renewed interest in FWI of hydrogen flames \cite{Mari2016, Zhao2022, DeNardi2024, Zhu2024, Chi2024}.
These studies primarily focus on either turbulent configurations or one-dimensional HOQ setups, where instabilities are inherently suppressed by the configuration.

While the effects of thermodiffusive instabilities in free flames and the FWI of hydrogen flames have been thoroughly studied, the FWI of thermodiffusive unstable flames has received little attention, except for a recent experimental study by Ojo et al.~\cite{Ojo2024}.
They conducted spatiotemporal surface temperature measurements resolving FWI of lean hydrogen/air flames using phosphor thermometry.
The measurements show that the wall temperatures exhibit alternating high- and low-temperature zones associated with finger-like flame structures, indicating that intrinsic combustion instabilities influence FWI. 
Still, it remains an open question to what extent the quenching characteristics are affected and what the governing physical phenomena are.
This detailed understanding is crucial for developing predictive combustion models that can be used in the design of novel combustion systems, for example, to determine the thermal loads on the combustor walls.

The present study aims to address this gap in literature regarding the FWI of thermodiffusively unstable flames, by conducting direct numerical simulations of laminar lean hydrogen/air flames in a two-dimensional HOQ configuration.
The main objectives are to:
\begin{itemize}[noitemsep]
    \item investigate the interaction of thermodiffusively unstable flames and combustor walls, and,
    \item characterize the effects of different cell size distributions of the instabilities on the quenching process.
\end{itemize}

The outline of the paper is as follows:
In Sec. \ref{sec:NumericalSetup}, the numerical setup for the simulation of two-dimensional HOQ of thermodiffusively unstable flames, as well as the configuration for generating the thermodiffusively unstable flames used as the initial condition for the HOQ, are detailed.
Furthermore, this section describes the numerical methods used in this study.
The thermodiffusively unstable freely-propagating flame is briefly characterized in Sec. \ref{sec:characterization_unquenched_FP_Flame}.
The interaction of the thermodiffusively unstable flame and a wall is analyzed and characterized in Sec. \ref{sec:2DHOQ}.
To obtain additional insights, the investigations are extended by setups that constrain the maximum cell size of the instabilities.
Finally, the main conclusions are presented in Sec. \ref{sec:Conclusions}.

\section{Numerical setup} \addvspace{10pt}
\label{sec:NumericalSetup}
In the following, the numerical configurations employed in this study are described before the numerical methods are presented.

\subsection{Configuration}
In this study a two-dimensional HOQ is investigated, with the setup described below.
An unconfined freely-propagating flame is used to generate the initial conditions, which is briefly described afterwards.

\subsubsection{Head-on quenching}

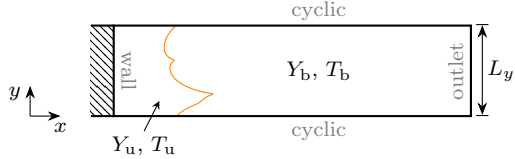
\begin{figure}[h!]
    \centering
    \begin{tikzpicture}
        \begin{axis}[
            at={(-1.6cm,-0.12cm)},
            axis lines=none, 
            clip mode=individual,
            width=2.35cm,
            height=3.03cm,
        ]
        
        \addplot[orange] table[col sep=comma, x index=0, y index=1] {data_for_figure1.csv};
        
        \end{axis}
    
        \draw[black, thick] (-2.2,0) rectangle (2.5,1.2);
        \node [] at (-1.8, -0.4) {\scriptsize $Y_{\mathrm{u}}$, $T_{\mathrm{u}}$};
        \draw[->] (-1.8, -0.2) -- (-1.6, 0.2);
        \node [] at (0.5, 0.6) {\scriptsize $Y_{\mathrm{b}}$, $T_{\mathrm{b}}$};
        \draw[thick](-2.5,0)--++(0.3,0);
        \draw[thick](-2.5,1.2)--++(0.3,0);
        \fill[pattern=north west lines] (-2.5,0) rectangle (-2.2, 1.2);

        \draw [-{Stealth}](-3.3,0) -- (-3.3,0.4);
        \draw [-{Stealth}](-3.3,0) -- (-2.9,0);
        \node [] at (-2.9, -0.15) {\scriptsize $x$};
        \node [] at (-3.5, 0.3) {\scriptsize $y$};
        \begin{scope}[>=Stealth]
            \draw [|<->|] (2.65, 0) -- (2.65, 1.22);
        \end{scope}
        \node [] at (2.9, 0.6) {\scriptsize $L_y$};

        \node[rotate=-90, color=gray] at (-2, 0.6) {\scriptsize wall};
        \node[rotate=90, color=gray] at (2.3, 0.6) {\scriptsize outlet};
        \node[color=gray] at (0.5, 1.4) {\scriptsize cyclic};
        \node[color=gray] at (0.5, -0.2) {\scriptsize cyclic};
    \end{tikzpicture}
    \caption{
        \footnotesize
        Schematic of the two-dimensional computational domain for the HOQ of the thermodiffusively unstable flames.
        The flame front is mapped from the initial simulations.
        The height of the domain in $y$-direction $L_y$ is varied.
    }
    \label{fig:configuration_FWI}
\end{figure}

\begin{table}[h!]
    \caption{\footnotesize
        Overview of the dimensions of the domain $L_x$ and $L_y$ in $x$ and $y$ direction for the different values of $\lambda$, which specifies the constraint of the domain in $y$ direction.
        num(HOQs) denotes the number of HOQ simulations performed for the respective value of $\lambda$.
    }
    \begin{tabular}{l|l|l|l}
    $\lambda$ & $L_y / \SI{}{mm}$                         & $L_x / \SI{}{mm}$   & num(HOQs)     \\ \hline
    2         & $2 \delta_{T}^0 = 1.22$   & $33.333 \delta_{T}^0 = 20.333$      & 1             \\
    4         & $4 \delta_{T}^0 = 2.44$   & $33.333 \delta_{T}^0 = 20.333$      & 1             \\
    8         & $8 \delta_{T}^0 = 4.88$   & $33.333 \delta_{T}^0 = 20.333$      & 50            \\
    16        & $16 \delta_{T}^0 = 9.76$  & $33.333 \delta_{T}^0 = 20.333$      & 25            \\
    32        & $32 \delta_{T}^0 = 19.52$ & $66.666 \delta_{T}^0 = 40.666$      & 10            \\
    64        & $64 \delta_{T}^0 = 39.04$ & $66.666 \delta_{T}^0 = 40.666$      & 10             \\
    100       & $100 \delta_{T}^0 = 61$   & $100 \delta_{T}^0 = 61$             & 4             \\
    \end{tabular}
\label{tab:parameters}
\end{table}

The two-dimensional computational domain for the two-dimensional HOQ of the thermodiffusively unstable flames is shown in Fig.~\ref{fig:configuration_FWI}.
In the streamwise direction ($x$), a wall is located on the left side, while an outlet is defined on the right side.
Periodic boundary conditions are enforced in the vertical direction at both the top and bottom of the domain.
The temperature of the wall corresponds to the temperature of the unburnt mixture ($T_{\mathrm{wall}} = T_{\mathrm{u}} = \SI{298}{\kelvin}$), a no-slip boundary condition is applied for the velocity and a zero flux boundary condition is specified for the species.
The zero flux boundary condition ensures that the mass flux of all species at the wall boundary remains zero by setting the total diffusion velocity at the wall in wall normal direction to zero (for details see section \ref{subsec:numerical_methods} on numerical methods).
At the outlet a zero-gradient boundary condition is used for the scalars and the velocity.
The simulations are initialized with temperature, velocity, and species profiles from initial simulations of thermodiffusive unstable hydrogen/air flames in the nonlinear regime with an equivalence ratio of $\varphi_{\mathrm{u}} = 0.4$, an unburned temperature of $T_\mathrm{u} = \SI{298}{K}$, and ambient pressure.
The corresponding setup for generating the initial condition is described in the following section.
The dimensions $L_y$ and $L_x$ of the domain, which are varied in this work, are presented in Table \ref{tab:parameters} and are based on established values from the literature \cite{Frouzakis2015, Berger2019}.
These dimensions are expressed as multiples of the thermal flame thickness $\delta_{T}^0$ of a laminar freely-propagating flame under the specified conditions of the unburnt mixture.
In the primary simulation, the domain height in the $y$-direction is set to $L_y = 100 \delta_T^0$, ensuring that no length scales of the intrinsic instabilities are suppressed \cite{Berger2019}.
This simulation is the main focus of the study.
The domains with smaller lateral dimensions (smaller values of $\lambda$) are selected to highlight the effects of instabilities at specific length scales.
The domain length in the streamwise direction, $L_x$, is deliberately set to be sufficiently large depending on the domain height, $L_y$, to ensure that all relevant structures of the instabilities are captured within the domain.
For $\lambda=100$, four HOQ simulations initialized from different time steps of the initial two-dimensional freely-propagating flames have been performed, for values of $\lambda < 100$ the number of HOQ simulations (num(HOQs)), is also given in Tab.~\ref{tab:parameters}.

A resolution of $20$ points per flame thickness is chosen, resulting in a uniform grid resolution of $\Delta_x = \Delta_y = \SI{30.5}{\micro\meter}$.

\subsubsection{Initial unsteady freely-propagating flame} 

\begin{figure}[h!]
    \centering
    \begin{tikzpicture}
        \draw[smooth, samples=200, domain=0:1.2, color=orange] plot({sin((\x)*5 r)*0.1+0.1}, \x);
        \draw[black, thick] (-2.2,0) rectangle (2.5,1.2);
        \node [] at (-1.2, 0.6) {\scriptsize $Y_{\mathrm{u}}$, $T_{\mathrm{u}}$};
        \node [] at (1.5, 0.6) {\scriptsize $Y_{\mathrm{b}}$, $T_{\mathrm{b}}$};
        \draw [-{Latex[round]}](-2.65,0.05) -- (-2.2,0.05);
        \draw [-{Latex[round]}](-2.65,0.27) -- (-2.2,0.27);
        \draw [-{Latex[round]}](-2.65,0.49) -- (-2.2,0.49);
        \draw [-{Latex[round]}](-2.65,0.71) -- (-2.2,0.71);
        \draw [-{Latex[round]}](-2.65,0.93) -- (-2.2,0.93);
        \draw [-{Latex[round]}](-2.65,1.15) -- (-2.2,1.15);
        \node [] at (-3.25, 0.93) {\scriptsize $u_{\mathrm{in}} \approx s_{\mathrm{c}}$};
        \draw [-{Stealth}](-3.3,0) -- (-3.3,0.4);
        \draw [-{Stealth}](-3.3,0) -- (-2.9,0);
        \node [] at (-2.9, -0.15) {\scriptsize $x$};
        \node [] at (-3.5, 0.3) {\scriptsize $y$};
        \begin{scope}[>=Stealth]
            \draw [>|-|<] (-0.17,0.1) -- (0.38,0.1);
            \draw [|<->|] (2.65, 0) -- (2.65, 1.22);
            \draw [|<->|] (-2.2, 1.35) -- (2.5, 1.35);
        \end{scope}
        \node [] at (0.1, -0.2) {\scriptsize $A_0$};
        \node [] at (2.9, 0.6) {\scriptsize $L_y$};
        \node [] at (-1.2, 1.55) {\scriptsize $L_x$};

        \node[rotate=-90, color=gray] at (-2, 0.6) {\scriptsize inlet};
        \node[rotate=90, color=gray] at (2.3, 0.6) {\scriptsize outlet};
        \node[color=gray] at (1, 1.55) {\scriptsize cyclic};
        \node[color=gray] at (1, -0.2) {\scriptsize cyclic};
    \end{tikzpicture}
    \caption{\footnotesize
    Schematic of the two-dimensional computational domain for generating the thermodiffusively unstable flames.
    The initial flame front consists of weakly perturbed one-dimensional unstretched flames.
    }
    \label{fig:configuration_initial}
\end{figure}
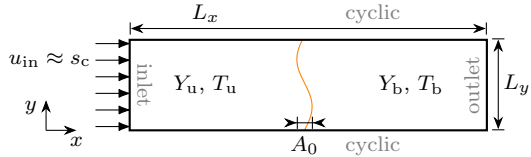

The two-dimensional computational domain for generating the initial conditions is shown in Fig.~\ref{fig:configuration_initial}.
Except for the boundary conditions at the inlet on the left side of the domain, the boundary conditions are similar to those of the HOQ.
For the velocity at the inlet, a boundary condition is employed that calculates the global consumption speed on the fly and utilizes the corresponding values from the last $n=100$ time steps to calculate a moving average for the velocity inlet condition, to stabilize the flame in the center of the domain.
The operating conditions are consistent with those of the HOQ ($\varphi_{\mathrm{u}} = 0.4$, $T_{\mathrm{u}} = \SI{298}{\kelvin}$, $p = \SI{1}{atm}$).

The simulations are initialized by one-dimensional freely-propagating unstretched flames that are aligned in $x$-direction and perturbed with a weak initial perturbation $A = A_0 \cos{\left(kx\right)}$, where $A_0 = 0.01 \delta_{T}^0$ is the perturbation amplitude.
The wavelength $\lambda \delta_{T}^0 = 2 \pi / k$ of the perturbation matches the height of the domain $L_y$ in each of the cases.
Simulations were run for over $100$ flame times $\tau_T^0$ to ensure that the flame propagates within the non-linear regime, the flame dynamics are statistically stationary and the time-averaged consumption speed remains constant.
The dimensions $L_y$ and $L_x$ of the domain, correspond to those of the respective HOQ simulation.

\subsection{Numerical methods}
\label{subsec:numerical_methods}
Simulations in this work have been performed with the open-source CFD library OpenFOAM \cite{Weller1998}, which solves the reactive compressible Navier-Stokes equations using the Finite Volume Method (FVM).
The reactive flow solver used for the simulations is an in-house solver derived from the standard OpenFOAM \textit{reactingFoam} solver \cite{Boettler2024, Lulic2023, Schneider2024, Schuh2024}.
The fluid is assumed to be an ideal gas and chemical reactions are modeled by directly solving the chemical source terms employing the detailed reaction mechanism of Li et al.~\cite{Li2004} that contains $9$ species and $19$ reactions.
Using the mixture-averaged diffusion approximation \cite{Kee2017}, the transport properties of the mixture are obtained by Cantera \cite{cantera}.
Following the mixture-averaged thermal diffusion model proposed by Chapman and Cowling \cite{Chapman1990}, molecular diffusion due to the Soret effect is also included \cite{Schlup2017}.
Hence, the species transport equation reads:
\begin{equation}
    \frac{\partial \rho Y_k}{\partial t} + \frac{\partial \rho u_i Y_k}{\partial x_i} + \frac{\partial}{\partial x_i} \left({\rho Y_{k}V_{k,i}^{\mathrm{total}}}\right) = \dot{\omega}_k \, ,
\end{equation}
with $V_{k,i}^{\mathrm{total}}$ being the total diffusion velocity of species $k$ in direction $i$, $\rho$ the density of the mixture, and $\dot{\omega}_k$ the source term of species $k$.
The diffusive flux $\rho Y_{k}V_{k,i}^{\mathrm{total}}$ is given by
\begin{equation}
     \rho Y_{k}V_{k,i}^{\mathrm{total}} = - \rho D_{k,\mathrm{mix}} \frac{\partial Y_k}{\partial x_i} - \rho Y_k D_{k,\mathrm{mix}} \frac{1}{\overline{W}}\frac{\partial \overline{W}}{\partial x_i} - \frac{D_{k,{\mathrm{therm}}}}{T}\frac{\partial T}{\partial x_i} - \rho Y_{k}V_{i}^{\mathrm{corr}} \, ,
\end{equation}
with the mixture-averaged species diffusion coefficients $D_{i,\mathrm{mix}}$, the molar mass of the mixture $\overline{W}$, and the thermal diffusion coefficient $D_{k,\mathrm{therm}}$.
Since the mixture-averaged diffusion model is a first-order approximation, a correction velocity $V^{\mathrm{corr}}_{i}$ is employed to ensure mass conservation and is defined as:
\begin{equation}
    \rho Y_k V_i^{\mathrm{corr}} = Y_k \sum_{k=1}^N \rho Y_k V_{k,i} = Y_k \sum_{k=1}^N \left( -\rho D_{k,\mathrm{mix}} \frac{\partial Y_k}{\partial x_i}  - \rho D_{k,\mathrm{mix}} Y_k \frac{1}{\overline{W}}\frac{\partial \overline{W}}{\partial x_i} - \frac{D_{k,{\mathrm{therm}}}}{T}\frac{\partial T}{\partial x_i}\right) \, .
\end{equation}
This leads to the following condition for the zero flux boundary condition for the species at the wall:
\begin{equation}
    \left. \rho Y_{k}V_{k,x}^{\mathrm{total}} \right|_{\mathrm{wall}} = \left.\left( - \rho D_{k,\mathrm{mix}} \frac{\partial Y_k}{\partial x} - \rho D_{k,\mathrm{mix}} Y_k \frac{1}{\overline{W}} \frac{\partial \overline{W}}{\partial x} - \frac{D_{k,\mathrm{therm}}}{T} \frac{\partial T}{\partial x} - \rho Y_{k}V_{x}^{\mathrm{corr}}\right) \right|_{\mathrm{wall}} \stackrel{!}{=} 0 \, .
\end{equation}
The enthalpy transport equation reads:
\begin{equation}
    \frac{\partial \rho h}{\partial t} + \frac{\partial \rho u_i h}{\partial x_i} = 
    \frac{\partial}{\partial x_i}\left(\frac{\lambda}{c_{\mathrm{p}}} \frac{\partial h}{\partial x_i}\right)
    - \sum_{k=1}^{N}\left(\frac{\partial}{\partial x_i} \left( \frac{\lambda}{c_{\mathrm{p}}} h_{k} \frac{\partial Y_k}{\partial x_i}\right) \right)
    - \sum_{k=1}^{N}\left(\frac{\partial}{\partial x_i}\left(h_{k} \rho Y_k V_{k,i}^{\mathrm{total}}\right)\right)
    + \frac{d p}{d t} \, ,
\end{equation}
with the enthalpy $h$, the enthalpy of the species $h_{k}$, the thermal conductivity of the mixture $\lambda$, and the specific heat capacity of the mixture at constant pressure $c_{\mathrm{p}}$.

An implicit second-order backward differentiation formula is employed for time advancement and second-order discretization schemes are employed in space.
The time integration of the chemical source terms employs a time-implicit backward difference method, implemented in the stiff ODE solver CVODE, which is part of the SUNDIALS suite \cite{hindmarsh2005sundials, gardner2022sundials}, using pyJac \cite{pyJac} to generate the source code for the chemical ODE system's right-hand side and Jacobian.
In addition, a dynamic load balancing for the chemical systems is employed \cite{Tekgl2021}.
The code is well validated and has been used in prior investigations \cite{Boettler2024, Lulic2023, Schneider2024, Schuh2024}.


\section{Characterization of the unstable hydrogen/air flames prior to quenching} \addvspace{10pt}
\label{sec:characterization_unquenched_FP_Flame}
For the characterization of the two-dimensional freely-propagating flame prior to quenching, the simulation domain, in which no length scales of the intrinsic instabilities are actively suppressed~\cite{Berger2019} (i.e. $L_y = \lambda \delta_T^0 = 100 \delta_T^0$), is analyzed.

\begin{figure}
    \centering
    \includegraphics[]{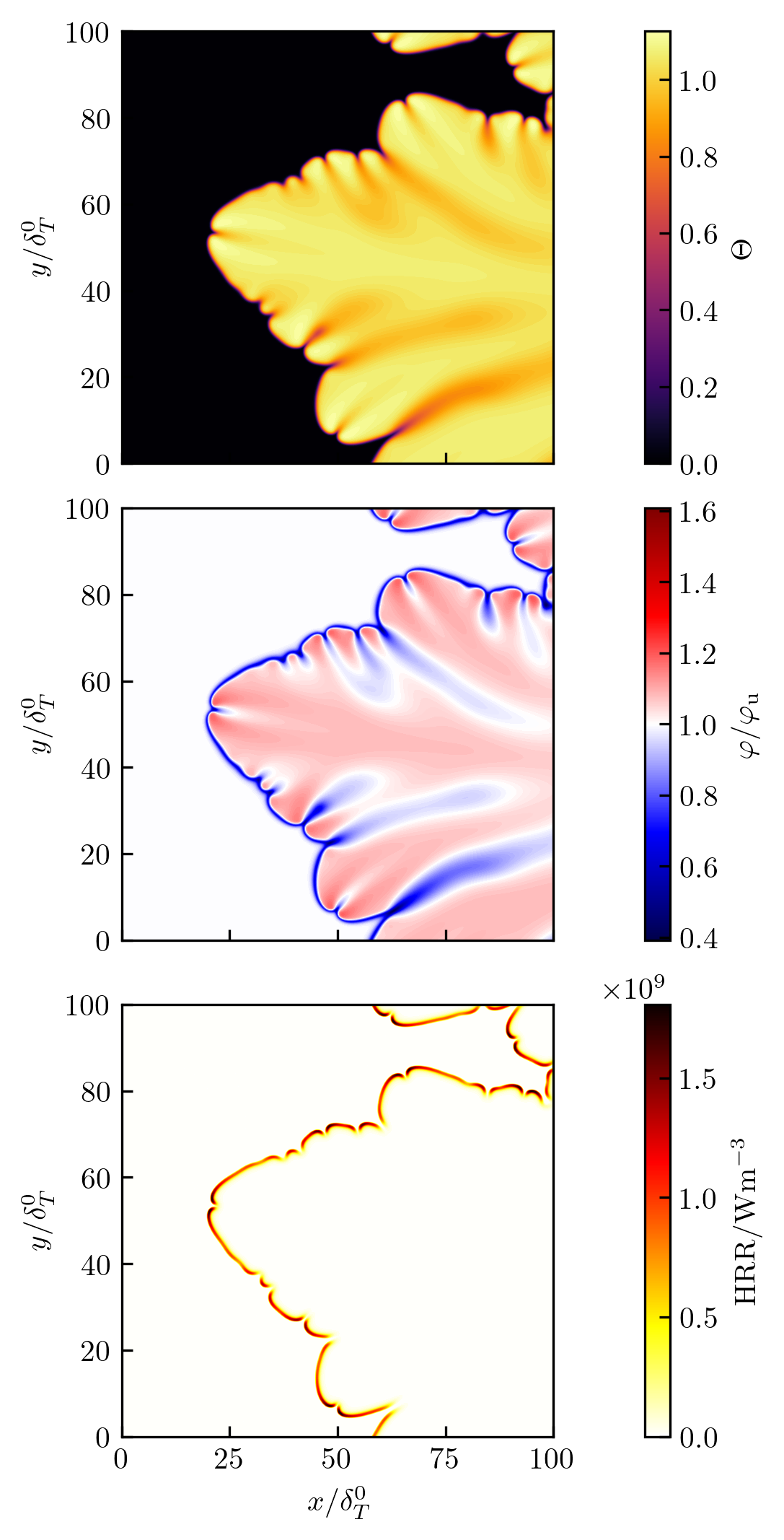}
    \caption{ \footnotesize
        Profiles of the normalized temperature $\Theta$ (top), the normalized local equivalence ratio $\varphi/\varphi_{\mathrm{u}}$ (middle) and the heat release rate $\mathrm{HRR}$ (bottom) over the numerical domain for $\lambda = 100 \delta_T^0$.
    }
    \label{fig:overview_unquenched_large_domain_part_I}
\end{figure}

To highlight the global flame structure of the thermodiffusively unstable flame, Fig.~\ref{fig:overview_unquenched_large_domain_part_I} shows a snapshot of the normalized temperature $\Theta = {\left(T - T_{\mathrm{u}}\right)}/{\left(T_{\mathrm{b}} - T_{\mathrm{u}}\right)}$, the normalized local equivalence ratio $\varphi/\varphi_{\mathrm{u}}$ and the heat release rate $\mathrm{HRR}$.
The normalization values correspond to the operation conditions prevailing in this study ($\varphi_{\mathrm{u}} = 0.4$, $T_{\mathrm{u}} = \SI{298}{K}$ and the flame temperature at equilibrium condition $T_{\mathrm{b}} = \SI{1422.06}{K}$). 

The observed flame displays a highly corrugated flame front with a wide range of length scales caused by the thermodiffusive instabilities.
In the center of the domain, so-called flame fingers can be identified which penetrate deep into the unburnt region, while smaller cells form on the sides of these fingers.
Further, regions with super-adiabatic temperatures ($\Theta > 1$) are observed, particularly in flame front segments curving convexly towards the unburnt mixture due to the locally enriched mixture ($\varphi / \varphi_{\mathrm{u}} > 1$). 
Conversely, in the wake of concavely curved segments, the opposite relationship holds ($\Theta < 1$ and $\varphi/\varphi_{\mathrm{u}} < 1$). 
These flame characteristics are in agreement with previous studies \cite{Altantzis2011, Altantzis2012, Berger2019} that also contain a more detailed analysis of these phenomena.

\begin{figure}[ht]
    \centering
    \includegraphics{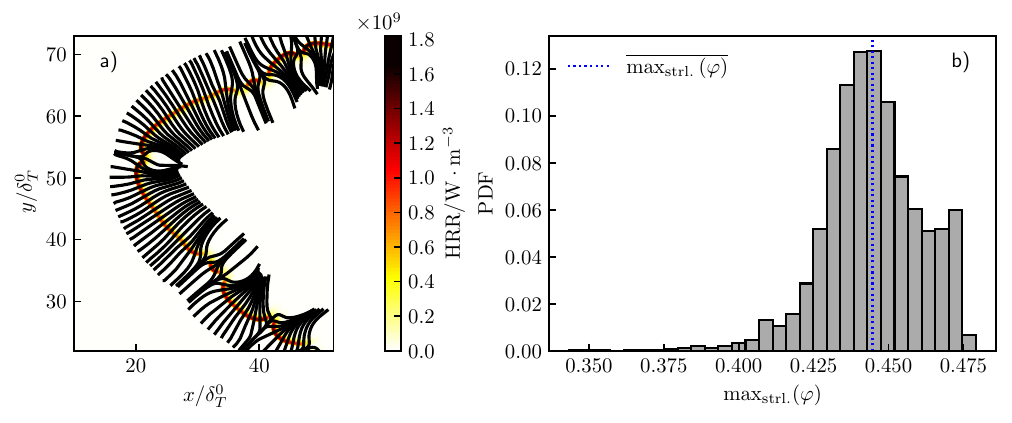}
    \caption{ \footnotesize
        a): Streamlines of the gradient of $Y_{\ce{H2}}$ in the area of the central flame finger (highlighted by the profile of the heat release rate).
        Every fifth streamline is shown.
        b): PDF of the maximum equivalence ratios along streamlines of the gradient of the $\ce{H2}$ mass fraction, $Y_{\ce{H2}}$, $\max_{\mathrm{streamline}}(\varphi)$, for $\lambda=100$.
        Data is extracted from $10$ timesteps for $1000$ streamlines per timestep.
    }
    \label{fig:phi_distribution_fp_flame_streamlines}
\end{figure}

To quantify the local mixture variation, which is evident from the global equivalence ratio profiles shown in Fig.~\ref{fig:overview_unquenched_large_domain_part_I} and serves as a marker for differential diffusion and preferential diffusion effects, $1000$ streamlines of the gradient of the mass fraction of hydrogen $Y_{\ce{H2}}$ (indicating the reaction progress) are extracted for $10$ different simulation time steps.
This follows the approach of Howarth and Aspden \cite{Howarth2022}, who defined tubes crossing the flame front along which they analyzed local flame quantities.
Figure \ref{fig:phi_distribution_fp_flame_streamlines} a) illustrates sample streamlines for a section of the numerical domain.
The maximum values\footnote{In one-dimensional freely-propagating hydrogen/air flames, the maximum equivalence ratio $\max(\varphi)$ along the flame corresponds to the equivalence ratio of the unburnt mixture, $\varphi_{\mathrm{u}}$.} of the equivalence ratio along the streamlines $\max_\mathrm{strl.}(\varphi)$ are utilized as indicators of mixture variation induced by the instabilities, shown as probability density function (PDF) in Fig.~\ref{fig:phi_distribution_fp_flame_streamlines} b).
The minimum value across all streamlines is $\min(\max_{\mathrm{strl.}}(\varphi)) = 0.34$, while the maximum value is $\max(\max_{\mathrm{strl.}}(\varphi)) = 0.48$.
The mean value, indicated by the dashed line, is $\overline{\max_{\mathrm{strl.}}(\varphi)} = 0.444$, which is $\approx 11\%$ larger than the equivalence ratio of the unburnt mixture.
This distribution of the equivalence ratio highlights the local mixture variation due to differential and preferential diffusion associated with the instabilities and will therefore be utilized as differential and preferential diffusion marker during the HOQ process in the following.

\begin{figure}[ht]
    \centering
    \includegraphics{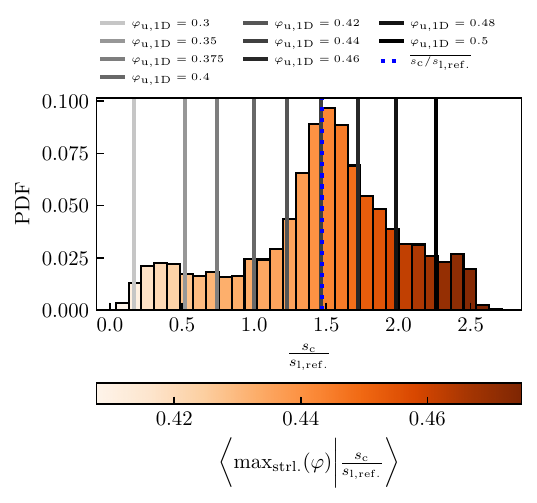}
    \caption{ \footnotesize
        PDF of the local consumption speed $s_{\mathrm{c}}$ (along streamlines of $Y_{\ce{H2}}$) normalized by the laminar flame speed of a one-dimensional freely-propagating flame at reference conditions $s_{\mathrm{l,ref.}}$ of the unburnt mixture colored in the conditional mean of the maximum equivalence ratios along the streamlines for each bin.
        The dashed blue line corresponds to the mean value.
        The grey lines correspond to one-dimensional freely-propagating flames at different values of the equivalence ratio of the unburnt mixture $\varphi_{\mathrm{u,1D}}$.
    }
    \label{fig:sc_distribuion_fp}
\end{figure}

In addition to the local mixture variation discussed above, the kinematic effects caused by the increase in flame surface area is another important phenomenon to consider in thermodiffusively unstable flames.
To separate the effects of local mixture variation from the kinematic effects, Fig.~\ref{fig:sc_distribuion_fp} shows a probability density function (PDF) of the normalized local consumption speed $s_{\mathrm{c}}/s_{\mathrm{l,ref.}}$.
Each bin is colored by the mean value of the maximum equivalence ratio along the streamlines ($\max_{\mathrm{streamline}}(\phi)$).
The reference conditions ($\mathrm{ref.}$) correspond to the conditions of the unburnt mixture ($\varphi = 0.4$, $T_{\mathrm{u}} = \SI{298}{\kelvin}$, $p=\SI{1.01325}{bar}$).
The consumption speed for each streamline is determined as
\begin{equation}
    s_{\mathrm{c}} = - \frac{1}{\rho_{\mathrm{u}} Y_{\ce{H2}\mathrm{,u}}}\int_0^{l_{\mathrm{strl.}}} \dot{\omega}_{\ce{H2}} \mathrm{d}l \quad ,
\end{equation}
where $l_{\mathrm{strl.}}$ is the length of the streamline, $l$ is the coordinate along the streamline and $\dot{\omega}_{\ce{H2}}$ is the fuel consumption rate.
An increased normalized local consumption speed with an increasing equivalence ratio is observed.
This trend can be well captured qualitatively by the one-dimensional freely-propagating flames at different values of the equivalence ratios $\varphi_{\mathrm{u,1D}}$.
The mean value $\overline{s_{\mathrm{c}}/s_{\mathrm{l,ref.}}}$ nearly perfectly corresponds to the respective value of a one-dimensional flame at a global equivalence ratio $\varphi_{\mathrm{u,1D}} = 0.44$.
Furthermore, this suggests, that the significantly higher average normalized global consumption speed of approximately $s_{\mathrm{c}}/s_{\mathrm{l}} = 3.7$ compared to the average normalized local consumption rate of about $\overline{s_{\mathrm{c}}/s_{\mathrm{l,ref.}}} = 1.5$ is due to an increase in flame surface area.
From these observations, in accordance with \cite{Howarth2022}, two main conclusions can be drawn:
\begin{itemize}[noitemsep]
    \item The variations in local flame consumption speed are primarily attributed to local mixture variation due to differential and preferential diffusion effects.
    \item The global flame consumption speed is additionally influenced by kinematic effects caused by an increase in flame surface area.
\end{itemize}
To analyze which of these effects most significantly influences the quenching process, after this characterization of the unstable flame prior to quenching, in the following the head-on quenching is analyzed.

\section{Analysis of the quenching process of the unstable flame} \addvspace{10pt}
\label{sec:2DHOQ}
The discussion of the thermodiffusively unstable freely propagating flame in the previous section reveals significant local mixture variations associated with the instabilities, as well as additional kinematic effects resulting from an increase in flame surface area.
The subsequent analysis examines the effects of thermodiffusive instabilities on the quenching process of lean premixed hydrogen/air flames, with particular emphasis on the distinct roles of local mixture variations and kinematic effects.
First, the quenching process is analyzed qualitatively, followed by a quantitative analysis.
The section concludes with an examination of the influence of the cell size distribution of the instabilities on the quenching process.

\subsection{Qualitative analysis of the quenching process}
\label{subsec:qualitative_analysis_qp}

\begin{figure}[ht]
    \centering
    \includegraphics[scale=0.8]{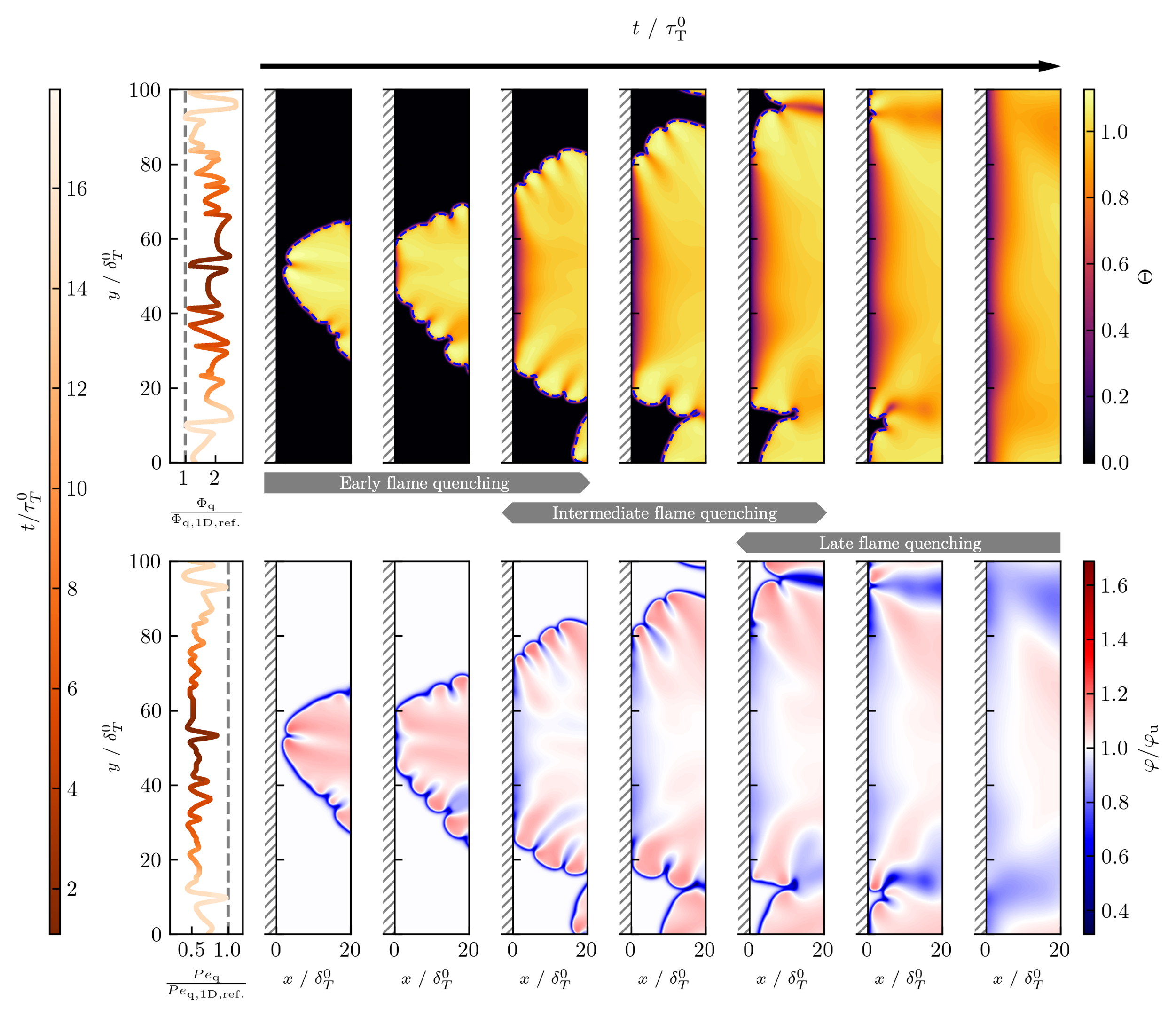}
    \caption{ \footnotesize
        Left: Quenching wall heat flux $\Phi_{\mathrm{q}}$ (top) and quenching Péclet number $Pe_{\mathrm{q}}$ (bottom) along the wall parallel coordinate $y$, normalized by the quenching wall heat flux $\Phi_{\mathrm{1D,ref.}}$ and the quenching Péclet number $Pe_{\mathrm{q,1D,ref.}}$ of the one-dimensional HOQ at reference conditions, respectively.
        The color scale indicates the quenching time step for the respective wall position, normalized by the characteristic flame time $\tau_T^0$ of the one-dimensional freely-propagating flame at reference conditions.
        Right: Normalized temperature $\Theta$ (top) and normalized local equivalence ratio $\varphi / \varphi_{\mathrm{u}}$ (bottom) over the numerical domain ($\lambda = 100$) for different time instances of the quenching process.
        The location of the wall is indicated by grey dashed lines.
        The direction of increasing time is indicated by the black arrow over the profiles, with different stages of the quenching process highlighted between the profiles.
        An animation of the time evolution corresponding to the figure can be found in the Supplementary Material.
    }
    \label{fig:T_phi_profile_2D_HOQ_largedomain}
\end{figure}

To visualize the quenching process, Fig.~\ref{fig:T_phi_profile_2D_HOQ_largedomain} (right) shows the time evolution of one of the HOQ simulations in the near-wall region with profiles of the normalized temperature $\Theta$ and normalized equivalence ratio $\varphi/\varphi_{\mathrm{u}}$.
The dashed blue line depicts the isoline $\mathrm{PV} = 0.9$, highlighting the flame front, which is defined following Howarth and Aspden \cite{Howarth2022} as $\mathrm{PV} = 1 - Y_{\ce{H2}}/Y_{\ce{H2}\mathrm{,u}}$.
Fig.~\ref{fig:T_phi_profile_2D_HOQ_largedomain} (top left) shows the quenching wall heat flux $\Phi_{\mathrm{q}}\left(y\right) = \max_t\left( \Phi\left(t,y\right) \right)$ along the wall, which corresponds to the temporal maximum of the wall heat flux, normalized by the quenching wall heat flux $\Phi_{\mathrm{q,1D,ref.}} = \max_t\left( \Phi_{\mathrm{1D,ref.}}(t) \right)$ for the one-dimensional HOQ at reference conditions.
The corresponding timestep is defined as the quenching time $t_{\mathrm{q}}$.
The dashed grey line highlights the quenching wall heat flux $\Phi_{\mathrm{q,1D,ref.}}$ of the one-dimensional HOQ at reference conditions.
The wall heat flux is calculated as
\begin{equation}
    \Phi = - \left. \lambda \cdot \frac{\partial T}{\partial x} \right|_{(x=0)} \, .
\end{equation}
Fig. \ref{fig:T_phi_profile_2D_HOQ_largedomain} (bottom left) shows the quenching Péclet number, which in this work is defined as the minimum of the Péclet number in time ($\min(Pe(t)$).
The Péclet number $Pe$ itself is defined as the wall distance of the flame (indicated by the maximum heat release rate along the wall normal coordinate $x$) normalized by the laminar flame thickness $\delta_T^0$ at reference conditions:
\begin{equation}
    Pe = \frac{x(\max(\mathrm{HRR}))}{\delta_T^0} \, .
\end{equation}
A time series animation corresponding to Fig.~\ref{fig:T_phi_profile_2D_HOQ_largedomain} can be found in the Supplementary Material.

The quenching process can be divided into three different stages, which are also denoted in Fig.~\ref{fig:T_phi_profile_2D_HOQ_largedomain}:
\begin{itemize}[noitemsep]
    \item \textbf{Early flame quenching}: 
    The flame finger at the center of the domain reaches the wall first, leading to frontal quenching in a HOQ manner, where the leading edges quench first, followed by the trailing edge.
    The leading and trailing edges are also discernible in the profiles of the normalized quenching wall heat flux and the normalized Péclet number.
    At the locations where the leading edges quench at the wall, the wall heat flux is high and the Péclet number is low; in contrast, trailing edges lead to the opposite characteristics.
    As a consequence of the quenching process, the cellular structures in the flame vanish locally.
    However, both in the wall-normal and wall-parallel directions the mixture variations in the flame persist even after the flame is extinguished at the wall.
    In areas farther from the wall, where the flame is still burning as indicated by the isoline of the progress variable, the structures of the instabilities remain and even new structures emerge.
    \item \textbf{Intermediate flame quenching}:
    After the leading part of the flame finger is extinguished, the thermodiffusively unstable flame propagates vertically upward and downward from the quenching location, locally quenching at different angles to the wall in a SWQ manner.
    During this process, additional cells/finger-like flame structures develop in the near-wall region.
    These structures subsequently quench at the wall, partially advancing beyond the main flame front and undergoing HOQ.
    High values of the quenching wall heat flux and low values of the quenching Péclet number can be observed in the regions of the individual cells, whereas the opposite is true for the areas between those cells.
    \item \textbf{Late flame quenching}:
    As time advances, multiple areas of the flame front merge (in this particular case in the lower and upper sections of the domain).
    This leads to locally leaner mixtures in the region of the combined flame front, which gradually approaches the wall.
    The quenching in these areas produces local wall heat flux and Péclet number values that closely resemble those observed in one-dimensional head-on quenching under equivalent unburnt mixture conditions ($\varphi = 0.4$).
    This completes the overall quenching process.
\end{itemize}

\begin{figure}
    \centering
    \includegraphics[scale=0.8]{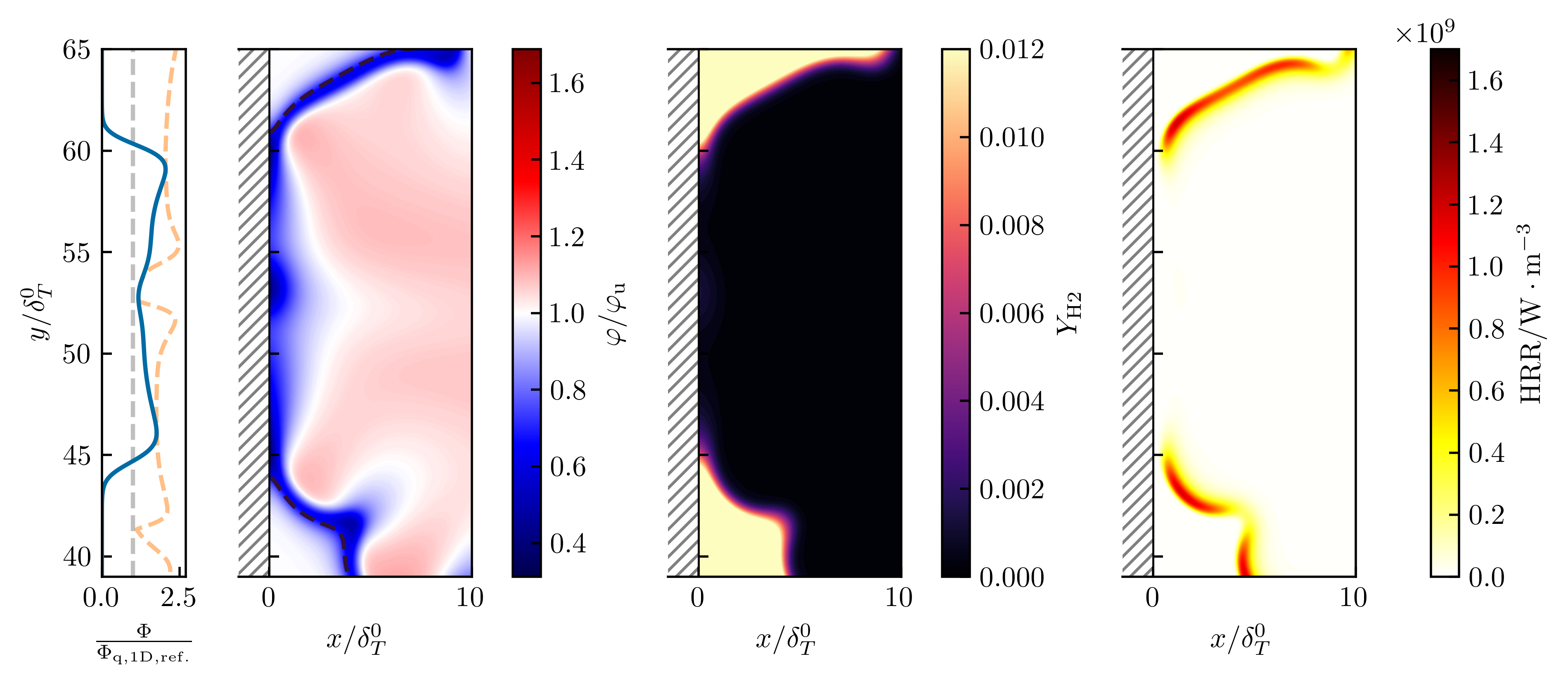}
    \caption{ \footnotesize
    Normalized wall heat flux $\Phi/\Phi_{\mathrm{q,1D,ref.}}$ (first column), normalized equivalence ratio $\varphi/\varphi_{\mathrm{u}}$ (second column), \ce{H2} mass fraction $Y_{\ce{H2}}$ (third column) and heat release rate HRR (fourth column) of the central flame finger in Fig.~\ref{fig:T_phi_profile_2D_HOQ_largedomain} quenching at the wall (single timestep, corresponding to the second time step shown in Fig.~\ref{fig:T_phi_profile_2D_HOQ_largedomain}).
    The section shown is part of the numerical domain as indicated by the dashed square in Fig.~\ref{fig:T_phi_profile_2D_HOQ_largedomain} for the corresponding time step.
    The dashed orange line in the first column corresponds to the normalized quenching wall heat flux $\Phi_{\mathrm{q}}/\Phi_{\mathrm{q,1D,ref.}}$, the dashed grey line corresponds to the quenching wall heat flux of the reference one-dimensional HOQ $\Phi_{\mathrm{q,1D,ref.}}$.
    The dashed black line in the second column highlights the isoline $PV = 1 - Y_{\ce{H2}} / Y_{\ce{H2}\mathrm{,u}}= 0.9$.
    An animated time series of this figure can be found in the Supplementary Material.
}
    \label{fig:phi_H2_whf_after_quenching_local}
\end{figure}

To emphasize the persistence of local mixture variations during and after quenching, as indicated in the analysis of the different stages of the quenching process, Figure \ref{fig:phi_H2_whf_after_quenching_local} highlights the local mixture variation $\varphi/\varphi_{\mathrm{u}}$, fuel mixture fraction $Y_{\ce{H2}}$ and heat release rate $\mathrm{HRR}$ for the central flame finger in Fig.~\ref{fig:T_phi_profile_2D_HOQ_largedomain} at the second timestep depicted in that figure.
An animated time series of Fig.~\ref{fig:phi_H2_whf_after_quenching_local} can be found in the Supplementary Material.
Even though the flame has already quenched locally (in the central region; $y / \delta_T^0 \approx 48 - 58$), as indicated by a vanishing $Y_{\ce{H2}}$, HRR and a decreasing wall heat flux, the local mixture variation, especially also in wall-parallel direction
\footnote{It should be noted that the local mixture variation in the wall-normal direction is partially due to the high diffusivity of molecular hydrogen, which diffuses away from the wall towards the flame front and can therefore be observed even in one-dimensional HOQs.}
, persists.
The original structure of the flame finger, characterized by its leading and trailing edges, can still be discerned from the profile of the normalized equivalence ratio even after quenching.
The time scale of diffusion is therefore larger than the time scale of quenching, indicating the importance of the local mixture variations on the quenching process.

To elucidate the implications of the thermodiffusive instabilities on the quenching process across all time steps, the profiles of the normalized quenching wall heat flux and the quenching Péclet number in Fig.~\ref{fig:T_phi_profile_2D_HOQ_largedomain} (left) are further examined.
In general, the quenching wall heat flux $\Phi_{\mathrm{q}}$ for nearly all values of $y$ is greater than that of the one-dimensional HOQ $\Phi_{\mathrm{q,1D,ref.}}$ under the same operating conditions.
In the areas where the flame is convexly curved (leading edges) in the near-wall region prior to quenching, the quenching wall heat flux is significantly larger than in the negatively curved areas (trailing edges).
This is evident from the flame finger at the center of the domain  (Fig.~\ref{fig:T_phi_profile_2D_HOQ_largedomain} (right)), where $\Phi_{\mathrm{q}}$ is significantly lower in the region with negative curvature (in its center) compared to the two leading points, which exhibit positive curvature.
A similar trend is observed for the quenching Péclet number, as $Pe_{\mathrm{q}}$ is lower than the one-dimensional HOQ value under identical conditions, $Pe_{\mathrm{q,1D,ref.}}$, in nearly all regions.
In the regions where $Pe_{\mathrm{q}}$ is low, the wall heat flux is high, and vice versa, indicating variations in the flame's behavior near the wall at different locations.
Consequently, the regions where the flame front is positively and negatively curved near the wall prior to quenching correspond well to higher and lower $Pe_{\mathrm{q}}$, respectively.
Accounting for curvature-related mixture fraction variations, regions with lower equivalence ratios (troughs, $\varphi/\varphi_{\mathrm{u}} < 1$) correlate with higher $Pe_{\mathrm{q}}$ and lower $\Phi_{\mathrm{q}}$, while regions with higher equivalence ratios (cusps, $\varphi/\varphi_{\mathrm{u}} > 1$) correlate with lower $Pe_{\mathrm{q}}$ and higher $\Phi_{\mathrm{q}}$.

The temporal evolution, along with the profiles of the normalized quenching wall heat flux and the normalized quenching Péclet number, clearly demonstrates that the quenching process influenced by thermodiffusive instabilities is substantially more complex than in a one-dimensional configuration.

\subsection{Quantitative analysis of the quenching process}
\label{subsec:quantitative_analysis_of_the_quenching_process}
\begin{figure}
    \centering
    \includegraphics[scale=0.9]{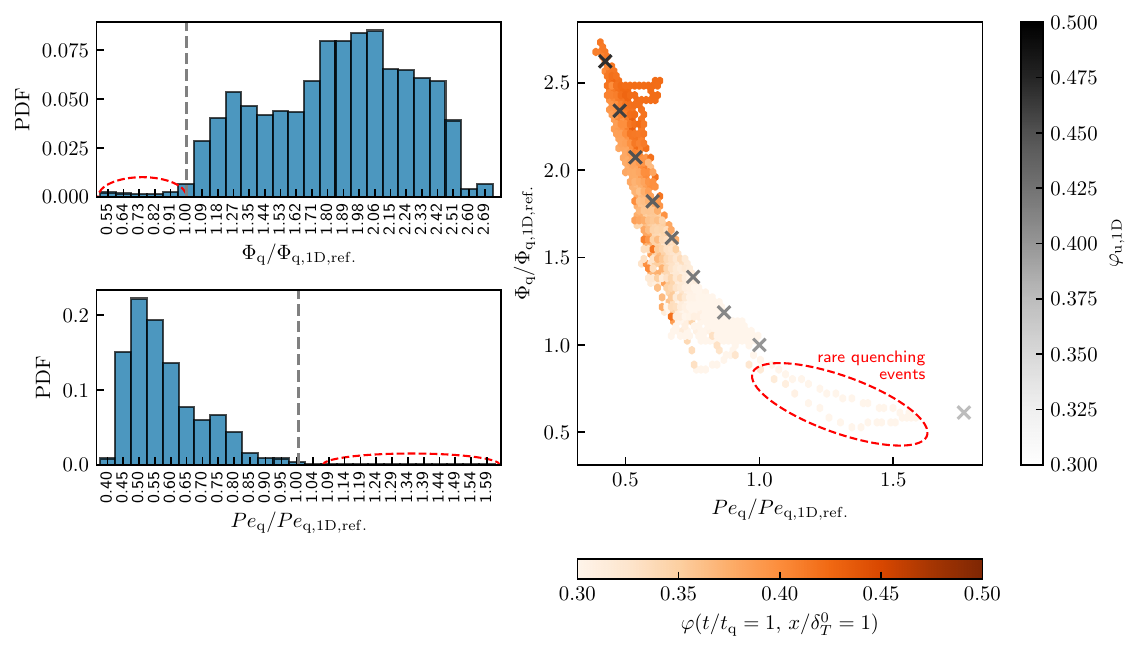}
    \caption{
    Left: PDF of the quenching wall heat flux $\Phi_{\mathrm{q}}$ (top) and Péclet number $Pe_{\mathrm{q}}$ (bottom) for $\lambda=100$, both normalized by the values of a one-dimensional HOQ at reference conditions $\Phi_{\mathrm{q,1D,ref.}}$ and $Pe_{\mathrm{q,1D,ref.}}$, respectively.
    Right: Normalized quenching wall heat flux $\Phi_{\mathrm{q}} / \Phi_{\mathrm{q,1D,ref.}}$ over the normalized quenching Péclet number $Pe_{\mathrm{q}} / Pe_{\mathrm{q,1D,ref.}}$ for $\lambda=100$.
    The colours indicate the local equivalence ratio at the distance of one thermal flame thickness ($\delta_T^0$ for $\varphi=0.4$) normal to the wall.
    The markers (`x`) colored in greyscale correspond to one-dimensional HOQs at different values of the unburnt mixture fraction $\varphi_{\mathrm{u,1D}}$.
    Rare quenching events are highlighted with a red ellipse.
    }
    \label{fig:PDFs_Peq_whf_largedomain}
\end{figure}

To quantify the influence of the thermodiffusive instabilities on the quenching process, Fig.~\ref{fig:PDFs_Peq_whf_largedomain} (left) shows the PDF of the normalized quenching wall heat flux, $\Phi_{\mathrm{q}}/\Phi_{\mathrm{q,1D,ref.}}$, and the normalized quenching Péclet number, ${Pe_{\mathrm{q}}}/{Pe_{\mathrm{q,1D,ref.}}}$.
The statistics shown are based on the $4$ HOQs performed for $\lambda = 100$ (see Tab.~\ref{tab:parameters}). 
In general, the unstable HOQ exhibits a lower quenching distance and a higher wall heat flux than the one-dimensional HOQ under similar operating conditions.
The exceptions are rare quenching events discussed in \ref{subsec:Appendix_lower_whf}. 

In addition to the PDF, Fig.~\ref{fig:PDFs_Peq_whf_largedomain} (right) shows the normalized quenching wall heat flux as a function of the normalized quenching Péclet number colored by the local equivalence ratio $\varphi$\footnote{The local equivalence ratio $\varphi$ is evaluated at one laminar flame thickness from the wall $x/\delta_T^0 = 1$ at the quenching time $t/t_{\mathrm{q}} = 1$}. 
For reference, a series of one-dimensional HOQs at varying equivalence ratios is also shown.
In the two-dimensional HOQ, a strong correlation between the local equivalence ratio, the Péclet number, and the wall heat flux is found.
This can be attributed to the decreasing flame thickness with increasing local equivalence ratio, which leads to a higher temperature gradient and a reduced quenching distance, resulting in an increased wall heat flux \cite{Poinsot1993}.
One of the main findings is that the observed trends in the two-dimensional HOQ are consistent with the one-dimensional references, suggesting that the global quantities characterizing the quenching process of the thermodiffusively unstable flame are primarily influenced by local mixture variations due to differential and preferential diffusion effects.

\begin{figure}
    \centering
    \includegraphics[]{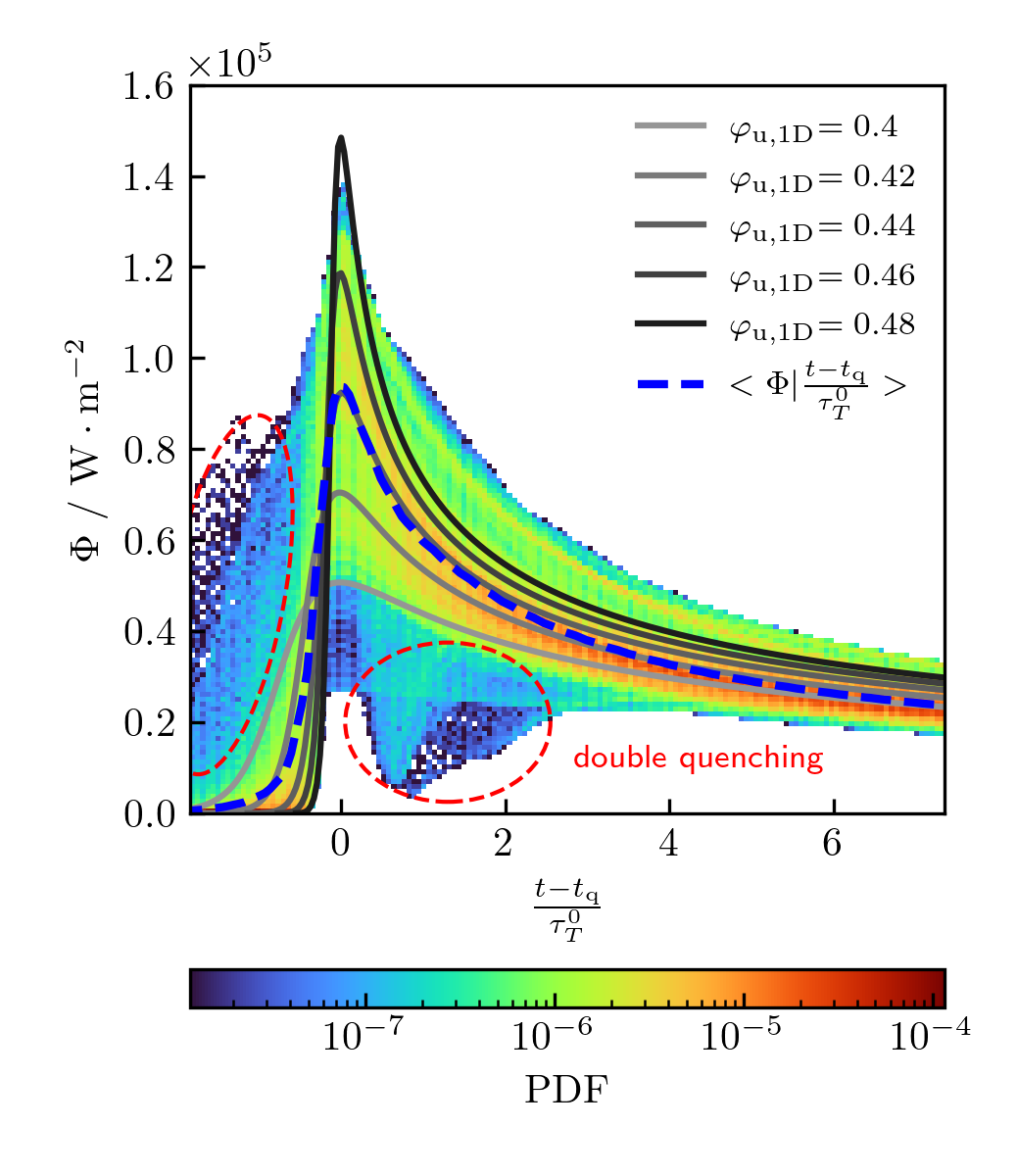}
    \caption{ \footnotesize
        PDF of the wall heat flux $\Phi$ over a relative normalized time ${(t-t_{\mathrm{q}})}/{\tau_{T}^0}$ (relative to the quenching time $t_{\mathrm{q}}$ and normalized by the characteristic flame time $\tau_{T}^0$ of a one-dimensional freely-propagating flame at reference conditions).
        The dashed blue line indicates the wall heat flux $\Phi$ conditioned on the normalized relative time ${(t-t_{\mathrm{q}})}/{\tau_{T}^0}$.
        The lines colored in different shades of grey correspond to one-dimensional freely-propagating flames at different values of the equivalence ratio $\varphi_{\mathrm{u,1D}}$.
    }
    \label{fig:whf_over_normalized_t_unconstrained_domain}
\end{figure}

In addition to the global quenching characteristics, Fig.~\ref{fig:whf_over_normalized_t_unconstrained_domain} shows the temporal evolution of the wall heat flux over a normalized relative time ${(t-t_{\mathrm{q}})}/{\tau_{T}^0}$.
The joint PDF corresponds to all the grid cells along the wall $y$ for the two-dimensional HOQs, while the solid lines correspond to one-dimensional HOQs at varying equivalence ratios.
Similar to the global quenching characteristics, the temporal evolution of quenching is well captured by mixture variations in the one-dimensional counterpart.
Specifically, both the upper and lower boundaries of the range containing the majority of data points (where PDF values exceed $1 \cdot 10^{-6}$) can be effectively enclosed by one-dimensional HOQs at varying equivalence ratio $\varphi_{\mathrm{u}}$.
Data points outside this range, highlighted by the red ellipses in the figure, can be attributed to rare double-quenching events that are discussed in \ref{subsec:Appendix_double_quenching}.

These findings are further emphasized by the averaged wall heat flux conditioned on the normalized relative time, $\langle\Phi|\frac{t-t_{\mathrm{q}}}{\tau_{T}^0}\rangle$ that almost perfectly aligns with a one-dimensional HOQ at $\varphi_\mathrm{u,1D} = 0.44$.
Notably, this equivalence ratio coincides with the condition under which the average consumption speed of the two-dimensional freely propagating flame (see Fig.~\ref{fig:sc_distribuion_fp}) matches that of a one-dimensional flame.
The quantification of the quenching process indicates that both the global flame characteristics of the unstable two-dimensional head-on quenching (HOQ) and the local quenching process, represented by the temporal evolution of the wall heat flux, are captured by a range of one-dimensional HOQs with varying equivalence ratio.

\subsection{Influence of instability cell size distribution on quenching}
\label{subsec:constrained_domains}
In this section, the effect of the instability cell size distribution on the quenching process is assessed.
A varying cell size distribution is achieved by restricting the numerical domain in the lateral ($y$) direction ($\lambda < 100$), thereby limiting the maximum combustion instability cell size.
For further details regarding the cell size distribution in freely-propagating flames for numerical domains of different sizes, see \cite{Berger2019}.

\begin{figure}
    \centering
    \includegraphics[scale=0.9]{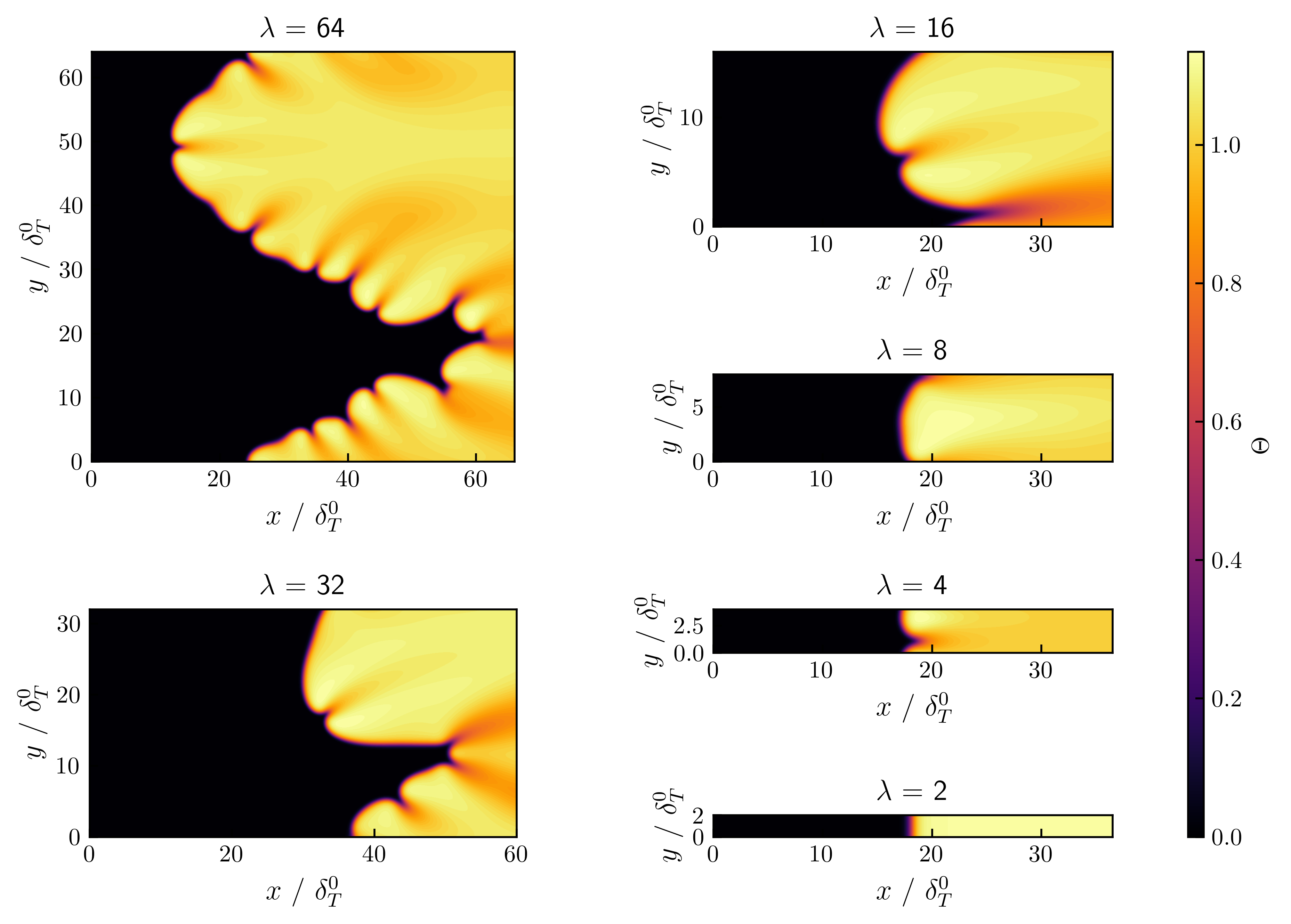}
    \caption{Profiles of the normalized temperature $\Theta$ over the numerical domain for the various constrained domains investigated ($\lambda < 100$).}
    \label{fig:constrained_domains}
\end{figure}

Figure \ref{fig:constrained_domains} illustrates the influence of cell size restriction on the instabilities showing the profiles of the normalized temperature $\Theta$ for two-dimensional freely-propagating flames for the various restricted domains investigated ($\lambda=2\,,4\,,8\,,16\,,32\,,64$; see Tab.~\ref{tab:parameters}).

For $\lambda=64$, large flame fingers extending deep into the unburnt region, with cusp formations along their sides, are observed.
The flame dynamics closely mirror those in an unrestricted domain, with the chaotic behavior typical of thermodiffusive instabilities.
As $\lambda$ decreases, both the lateral motion and the flame front's extent in the $x$-direction diminish, with the cusps along the edges of the flame fingers disappearing first, followed by the gradual disappearance of the flame fingers themselves.
Therefore, as indicated in \cite{Berger2019}, with decreasing $\lambda$ the cell size distribution narrows and large scale structures disappear.
For $\lambda=4$, only a single cell is formed that propagates uniformly and no lateral movement can be identified.
Shifting the initial displacement in $y$ direction does not lead to a different cell formation either.
Finally, for $\lambda=2$, the extent of the domain in the lateral direction is too small for thermodiffusive instabilities to develop and the flame propagation is similar to a one-dimensional freely-propagating flame.
The numbers of HOQ simulations, num(HOQs), conducted for the different values of $\lambda$ are given in Tab.~\ref{tab:parameters}.

\begin{figure}
    \centering
    \includegraphics[scale=0.8]{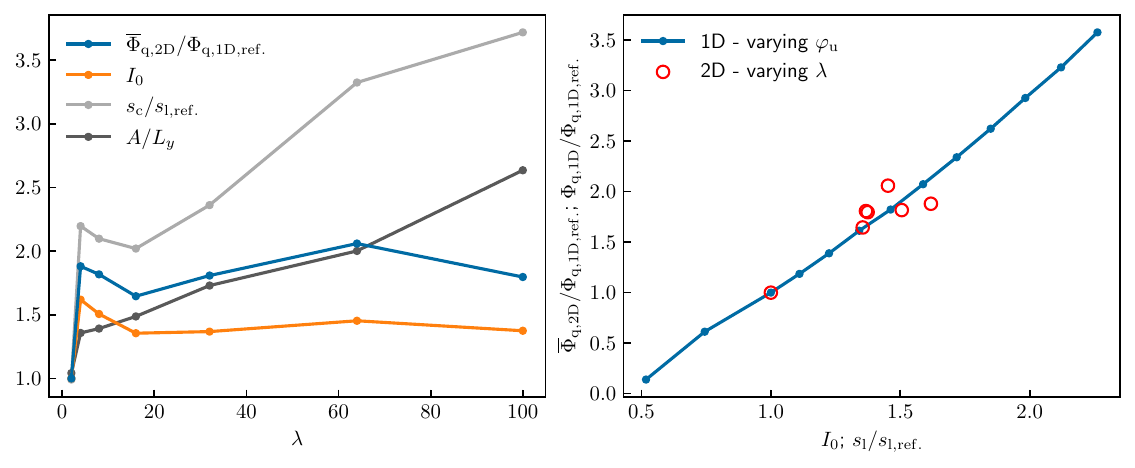}
    \caption{ \footnotesize
        Left: Normalized mean quenching wall heat flux ${\overline{\Phi}_{\mathrm{q,2D}}}/{\Phi_{\mathrm{q,1D,ref.}}}$, reactivity factor $I_0$, normalized flame surface area, $A/L_y$, and normalized consumption speed $s_{\mathrm{c}}/s_{\mathrm{l,ref.}}$ for different domain widths $L_y =  \lambda \delta_T^0$.
        Right: Normalized mean quenching wall heat flux ${\overline{\Phi}_{\mathrm{q,2D}}}/{\Phi_{\mathrm{q,1D,ref.}}}$ (values are extracted from the two-dimensional HOQs) versus the reactivity factor $I_0$ (values are extracted from the two-dimensional freely-propagating flames) for the different domain widths; normalized quenching wall heat flux ${\Phi_{\mathrm{q,1D}}}/{\Phi_{\mathrm{q,1D,ref.}}}$ (values are extracted from the one-dimensional HOQs) versus the normalized laminar flame speed $s_{\mathrm{l}} / s_{\mathrm{l,ref.}}$ for varying a equivalence ratio $\varphi_{\mathrm{u}}$ in the range $\varphi_{\mathrm{u}} = 0.3-0.5$.
    }
    \label{fig:whf_I0_over_lambda}
\end{figure}

Figure \ref{fig:whf_I0_over_lambda} (left) shows the reactivity factor $I_0$, the normalized surface area $A/L_y$ and the normalized consumption speed $s_{\mathrm{c}} / s_{\mathrm{l,ref.}}$ of the freely-propagating flames approaching the wall as a function of the different domain widths ($\sim \lambda$). 
The reactivity factor $I_0$ is determined from the normalized consumption speed $s_{\mathrm{c}} / s_{\mathrm{l,ref.}}$ and the normalized surface area $A / L_{y}$ (normalized by the inlet area):
\begin{equation}
    I_0 = \frac{s_{\mathrm{c}}}{s_{\mathrm{l,ref.}}} \bigg/ \frac{A}{L_y} \quad .
\end{equation}
In two dimensions, the flame surface is a line and its surface area $A$ is determined by the length of the isoline $\mathrm{PV} = 0.9$.
In addition to the properties of the freely propagating flame, the normalized mean quenching wall heat flux, ${\overline{\Phi}_{\mathrm{q,2D}}}/{\Phi_{\mathrm{q,1D,ref.}}}$ is also depicted as a quenching characteristic.

The normalized mean quenching wall heat flux shows no discernible correlation between the normalized consumption speed or flame surface area, indicating that the kinematics of the flame approaching the wall have no major impact on the subsequent quenching process.
However, the mean wall heat flux and the reactivity factor $I_0$ are strongly correlated, suggesting that the flame's reactivity prior to quenching significantly influences the subsequent quenching process.
Notably, the peak and subsequent decrease in reactivity observed for smaller domains with increasing $\lambda$ are reflected in the peak and subsequent decrease in the mean normalized quenching wall heat flux.
Thus, the objective is to identify the underlying cause of the observed change in flame reactivity.

To further elaborate on this finding, Fig.~\ref{fig:whf_I0_over_lambda} (right) displays the mean normalized quenching wall heat flux over the reactivity factor for the different domain sizes for the two-dimensional HOQs (red markers) and the quenching wall heat flux for one-dimensional HOQs at different equivalence ratios $\varphi_{\mathrm{u}}$ (blue line), over the normalized laminar flame speed, $s_l / s_{\mathrm{l,ref.}}$. 
The unstable flames approaching the wall at different domain sizes nearly perfectly align with the one-dimensional counterparts at varying equivalence ratio, again, indicating that the flame reactivity is primarily linked to local mixture variations associated with thermodiffusive instabilities.
Furthermore, Fig.~\ref{fig:whf_I0_over_lambda} (right) supports the conclusion that the intrinsic instabilities mainly influence the FWI through local mixture variation, as the values of the average wall heat flux for different domain sizes in lateral direction correspond well with those of one-dimensional HOQs at different equivalence ratios.

In summary, the above findings suggest that kinematic effects, such as an increased flame surface area, have a minor impact on key FWI quantities.
The dominant factor affecting flame quenching is the flame's reactivity before quenching, which is largely governed by local mixture variations due to differential and preferential diffusion effects.

\section{Conclusions} \addvspace{10pt}
\label{sec:Conclusions}

The present study investigates the interaction of thermodiffusive unstable hydrogen/air flames with isothermal walls in a two-dimensional head-on quenching configuration using direct numerical simulations.
In the first step, the freely propagating flame approaching the combustor wall is characterized.
Using a local streamline analysis, a strong correlation is observed between local mixture variation as a marker for differential and preferential diffusion effects and the local consumption speed of the flame.
Comparisons with one-dimensional flames at varying equivalence ratios show, that the local flame speed distribution of the unstable flame can be well captured and therefore local mixture variation must be considered to accurately capture the effects of intrinsic combustion instabilities on the local flame propagation.

Subsequently, the interaction of unstable hydrogen/air flames with the wall is evaluated.
Averaged over the entire height of the wall, the unstable flame exhibits a higher wall heat flux and a shorter quenching distance compared to a one-dimensional head-on quenching under the same operating conditions.
However, locally, the observed heat fluxes correspond well with those of one-dimensional head-on quenching flames at the same local equivalence ratios, displaying the same wall heat fluxes and quenching distances.
These findings are further supported by a subsequent analysis of head-on quenching configurations at varying domain sizes, allowing the assessment of the effect of the instability cell size distribution on the quenching process.
The results show that the variation in the flame's reactivity, caused by local mixture variation due to differential and preferential diffusion, and the quenching wall heat flux are correlated, i.e., increased reactivity results in a higher wall heat flux and vice versa.
This relationship is consistent with the correlation between the laminar burning velocity and the quenching wall heat flux in one-dimensional head-on quenching simulations across different equivalence ratios, quantifying the significant role of local mixture variation associated with the instabilities in flame-wall interaction.

In conclusion, our findings suggest that the unstable flame-wall interactions can be represented by an ensemble of one-dimensional head-on quenchings at different equivalence ratios.
Future research will address the incorporation of these effects into flamelet manifolds.
In this context, it should be emphasized that the local mixture distribution must be resolved in each modeling approach.
In part II of this study, the effects of instabilities on the flame-wall interaction are investigated across a range of different pressures $p$, equivalence ratios $\varphi$, and unburnt gas temperatures $T_{\mathrm{u}}$ to assess whether the findings can be transferred to different operating conditions.

\newpage

\section*{Disclosure statement}
The authors declare that they have no known competing financial interests or personal relationships that could have appeared to influence the work reported in this paper.

\section*{Acknowledgements}
This work has been funded by the Deutsche Forschungsgemeinschaft (DFG, German Research Foundation) – Project Number 523792378 - SPP 2419.
The simulations were performed on the Lichtenberg high-performance computer at TU Darmstadt.

\newpage

\appendix

\section{Rare quenching events}
\label{sec:AppendixA}

\subsection{Lower quenching wall heat flux}
\label{subsec:Appendix_lower_whf}

\begin{figure}[H]
    \centering
    \includegraphics[scale=0.8]{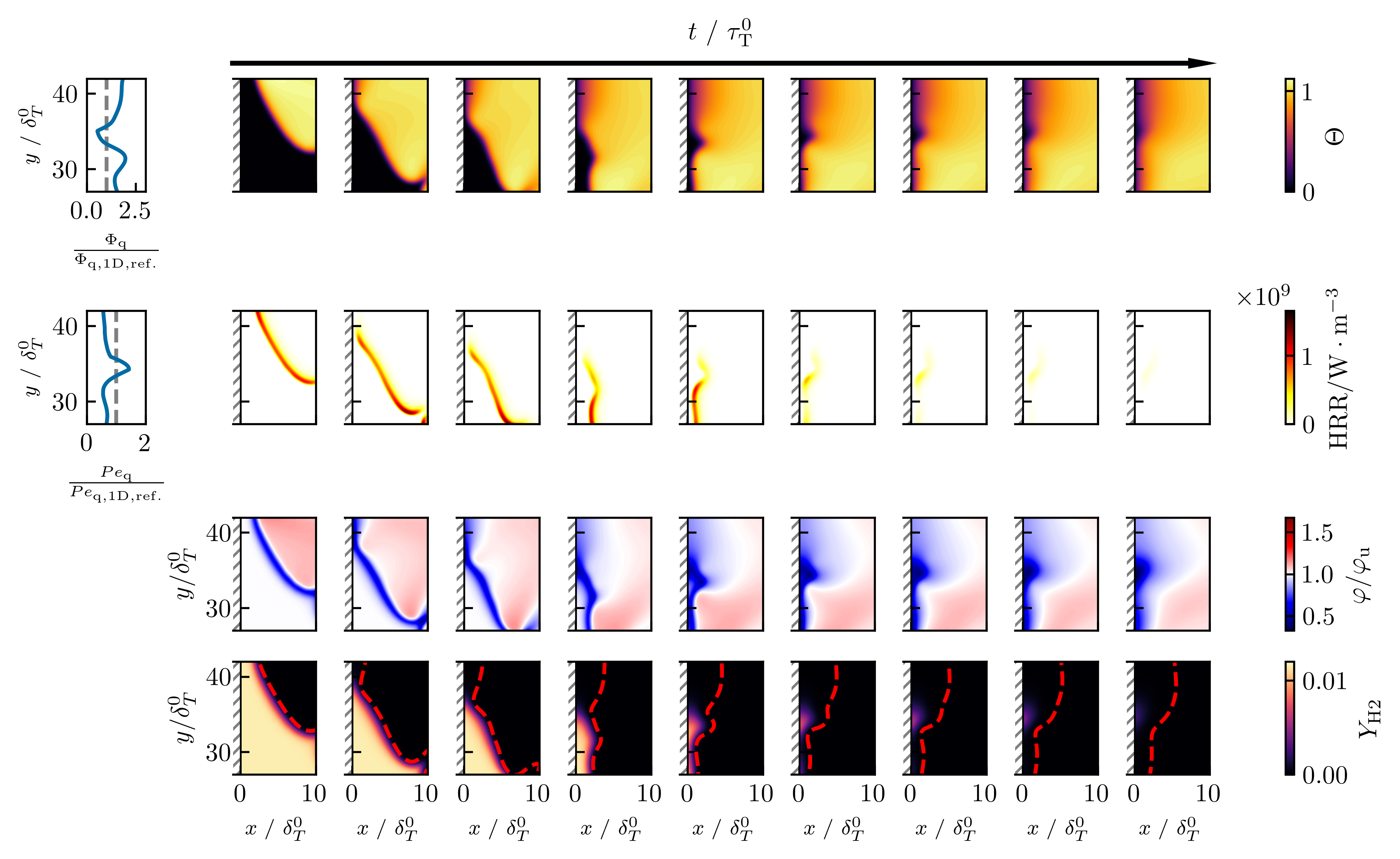}
    \caption{
        Temporal evolution of the normalized temperature $\Theta$ profiles (first row, right),
        heat release rate (second row, right), normalized equivalence ratio $\varphi / \varphi_{\mathrm{u}}$ (third row) and fuel mass fraction $Y_{\ce{H2}}$ (fourth row) over a subsection of the domain for a HOQ with $\lambda = 100$.
        The dashed red line in the fourth row depicts the isoline $\Theta = 0.75$.
        The normalized quenching wall heat flux ${\Phi_{\mathrm{q}}}/{\Phi_{\mathrm{q,1D,ref.}}}$ (first row, left) and the normalized quenching Péclet number $Pe_{\mathrm{q}} / Pe_{\mathrm{q,1D,ref.}}$ (second row, left) are shown for the corresponding subsection of the wall.
        The dashed grey lines highlight the values of the one-dimensional HOQ at $\varphi_{\mathrm{u}} = 0.4$.    
        An animation corresponding to this figure is provided in the Supplementary Material.
    }
    \label{fig:whf_smaller_than_1D}
\end{figure}

The small number of quenching events observed in Sec. \ref{subsec:quantitative_analysis_of_the_quenching_process}, where the quenching Péclet number is higher and the quenching wall heat flux is lower than for the reference one-dimensional HOQ ($\varphi_{\mathrm{u}} = 0.4$), are discussed in more detail based on a time series of a subsection of the numerical domain for one of the HOQ instances, shown in Fig.~\ref{fig:whf_smaller_than_1D}.
The flame front, which is indicated by the profile of the heat release rate, first progresses from the top right to the bottom left of the section.
As it reaches the wall in the upper left corner, it initially burns downwards along the wall and gradually quenches in SWQ manner. 
The local consumption speed near the wall continues to decrease due to the heat losses towards the wall until it completely extinguishes locally (directly on the wall).
Simultaneously, in the lower part of the observed region the flame propagates towards the wall, as the flame is not yet influenced by the wall.
This propagation is not perfectly perpendicular to the wall, as the leading edge also exhibits movement in the positive $y$ direction.
Although there is still unburnt fuel present in the region very close to the wall (as indicated by the profile of $Y_{\ce{H2}}$), this branch of the flame also locally quenches due to heat losses when it approaches very close to the wall.
Consequently, a very fuel-lean pocket remains.
The flame, significantly weakened by heat losses, is not able to propagate into this region outside the flammability limits.
Therefore, at the position of this pocket, the wall heat flux is low and the quenching Péclet number is high (see the corresponding profiles in Fig.~\ref{fig:whf_smaller_than_1D}).

\subsection{Double quenching}
\label{subsec:Appendix_double_quenching}

\begin{figure}
    \centering
    \includegraphics[]{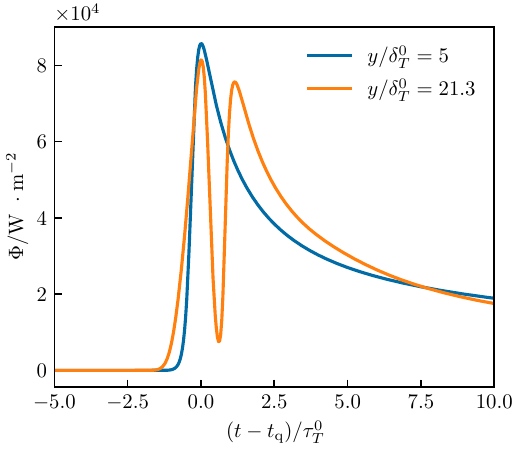}
    \caption{Wall heat flux $\Phi$ over a relative normalized time $\frac{t-t_{\mathrm{q}}}{\tau_{T}^0}$ (relative to the quenching time $t_{\mathrm{q}}$ and normalized by the flame time $\tau_{T}^0$ of a 1D freely-propagating flame at reference conditions) for two different points at the wall ($y/\delta_T^0=5$ and $y/\delta_T^0=21.3$).}
    \label{fig:whf_over_t_comparison_largeDomain}
\end{figure}

To further investigate the rare quenching events observed in the temporal profiles of the wall heat fluxes in Sec. \ref{subsec:qualitative_analysis_qp}, an analysis of the wall heat flux profiles for different wall positions ($y$) provides two distinct curve shapes, as illustrated in Fig.~\ref{fig:whf_over_t_comparison_largeDomain}:
\begin{itemize}
    \item Standard behaviour: Profiles (blue line as an example), which in their evolution (increasing wall heat flux, quenching and slow decrease due to cooling of the exhaust gas) resemble the usual one-dimensional HOQ like behaviour.
    This profile can be found for most locations along the wall.
    \item Double quenching: Profiles (orange line as an example), which show a peak in the wall heat flux, followed by a fast decrease, followed by another peak, after which the wall heat flux slowly decreases.
    This double quenching event only occurs at a few locations of the wall $y$, whereby the minimum between the two wall heat flux peaks and the magnitude of the peaks are pronounced to different extends.
\end{itemize}

\begin{figure}[ht]
    \centering
    \includegraphics[scale=0.9]{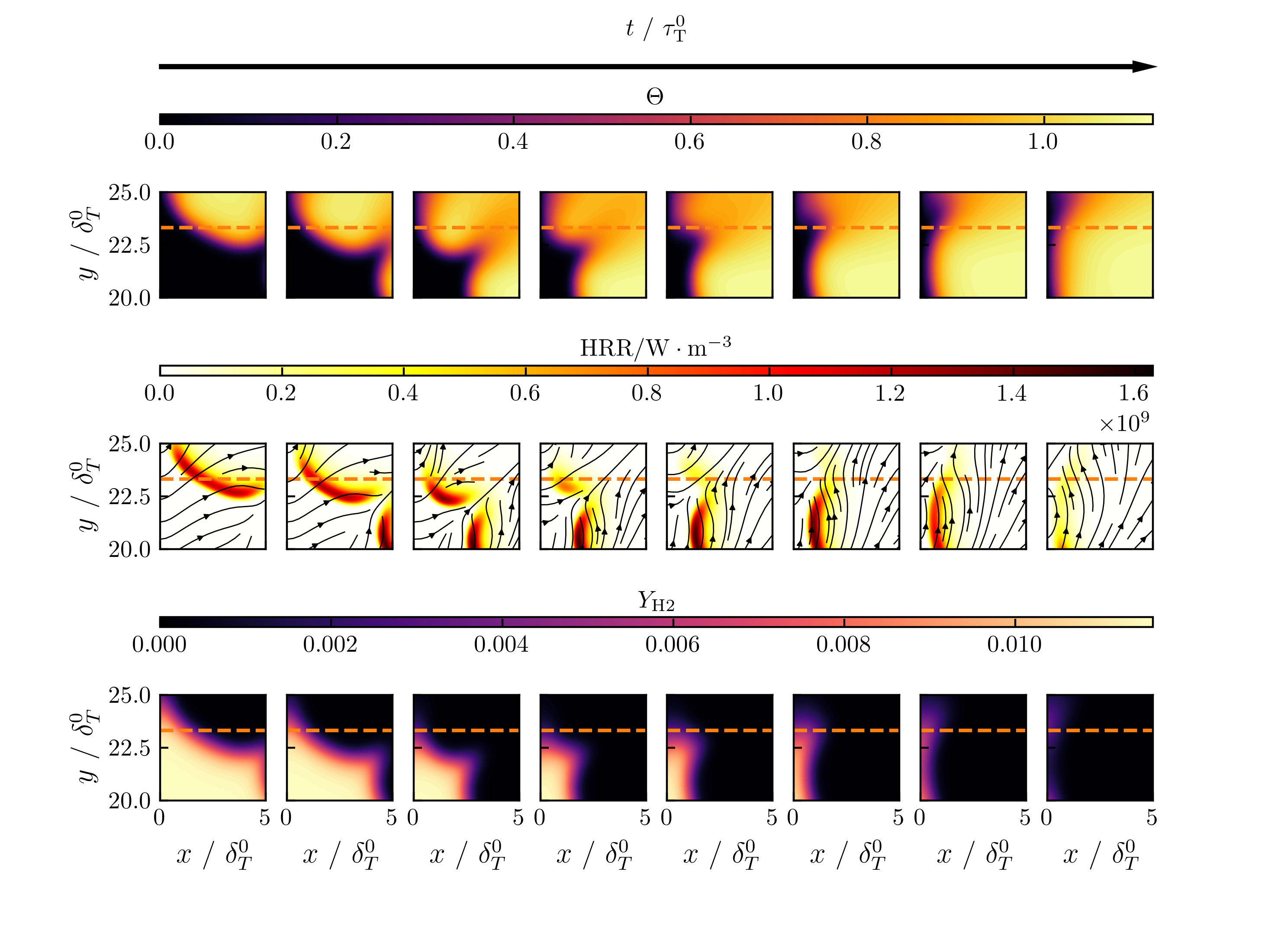}
    \caption{
        Multiple time instances of the normalized temperature $\Theta$ (top), heat release rate $\mathrm{HRR}$ (middle) and fuel mass fraction $Y_{\ce{H2}}$ in the near-wall area over a small region of the numerical domain.
        The black arrows in the middle row correspond to streamlines of the gradient of the fuel mass fraction ($\partial Y_{\ce{H2}} / \partial x_i$).
        The dashed orange line corresponds to the position of the orange line in Fig.~\ref{fig:whf_over_t_comparison_largeDomain} ($y/\delta_T^0=21.3$).
    }
    \label{fig:closeup_double_quenching_large_domain}
\end{figure}
To investigate the cause of this rare quenching event in more detail, Fig.~\ref{fig:closeup_double_quenching_large_domain} focuses on one of the double quenching events occurring in the near-wall region within a subsection of the numerical domain.
The dashed orange line correspond to the location of the orange line in Fig.~\ref{fig:whf_over_t_comparison_largeDomain} in the spatial domain ($y/\delta_T^0=21.3$).
From the top row of Fig.~\ref{fig:closeup_double_quenching_large_domain} it becomes evident, that the double quenching events can occur at the tips of the flame fingers, where the flame is concavely curved (towards the unburnt gases).
In this specific case, the upper part of the flame finger interacts with the wall leading to a high wall heat flux and therefore quenching at the wall as indicated by the heat realease rate (HRR) profiles in the middle row.
Since there still is unburnt molecular hydrogen (fuel) below the orange line, it not only diffuses in the direction normal to the wall (towards the flame front), but also partly in the direction parallel to the wall, as illustrated by the streamlines of $\frac{\partial Y_{\ce{H2}}}{\partial x_i}$ in the middle row.
This can also be attributed to the high diffusivity of the molecular hydrogen. 
The flame subsequently burns diagonally from below location of the orange line towards it as the molecular hydrogen diffuses in that direction, resulting in the second quenching event.

It is evident that the rare quenching events leading to lower wall heat fluxes and double quenching are quite similar. 
In both cases, the flame front initially quenches locally.
In the second case another branch of the flame continues to consume the fuel next to the quenching location, resulting in a second peak in the wall heat flux profile.
In contrast, the first case involves the formation of a fuel-lean pocket (outside the flammability limits), which locally leads to lower wall heat fluxes.

\newpage

\bibliographystyle{unsrtnat_mod}
\bibliography{publication.bib}

\begin{thebibliography}{47}
\providecommand{\natexlab}[1]{#1}
\providecommand{\url}[1]{\texttt{#1}}
\expandafter\ifx\csname urlstyle\endcsname\relax
  \providecommand{\doi}[1]{doi: #1}\else
  \providecommand{\doi}{doi: \begingroup \urlstyle{rm}\Url}\fi

\bibitem[Dreizler et~al.(2021)Dreizler, Pitsch, Scherer, Schulz, and
  Janicka]{Dreizler2021}
A.~Dreizler, H.~Pitsch, V.~Scherer, C.~Schulz, and J.~Janicka.
\newblock The role of combustion science and technology in low and zero impact
  energy transformation processes.
\newblock \emph{Appl. Energy Combust. Sci.}, 7:\penalty0 100040, September
  2021.
\newblock ISSN 2666-352X.

\bibitem[Verhelst and Wallner(2009)]{Verhelst2009}
S.~Verhelst and T.~Wallner.
\newblock Hydrogen-fueled internal combustion engines.
\newblock \emph{Prog. Energy Combust. Sci.}, 35:\penalty0 490–527, December
  2009.
\newblock ISSN 0360-1285.

\bibitem[Correa(1993)]{Correa1993}
S.~M. Correa.
\newblock A review of {NO$_{x}$} formation under gas-turbine combustion
  conditions.
\newblock \emph{Combust. Sci. Technol.}, 87:\penalty0 329–362, January 1993.
\newblock ISSN 1563-521X.

\bibitem[Kwon et~al.(2002)Kwon, Rozenchan, and Law]{Kwon2002}
O.~Kwon, G.~Rozenchan, and C.~Law.
\newblock Cellular instabilities and self-acceleration of outwardly propagating
  spherical flames.
\newblock \emph{Proc. Combust. Inst.}, 29:\penalty0 1775–1783, January 2002.
\newblock ISSN 1540-7489.

\bibitem[Bauwens et~al.(2017)Bauwens, Bergthorson, and Dorofeev]{Bauwens2017}
C.~Bauwens, J.~Bergthorson, and S.~Dorofeev.
\newblock Experimental investigation of spherical-flame acceleration in lean
  hydrogen-air mixtures.
\newblock \emph{Int. J. Hydrogen Energ.}, 42:\penalty0 7691–7697, March 2017.
\newblock ISSN 0360-3199.

\bibitem[Fernández-Galisteo et~al.(2018)Fernández-Galisteo, Kurdyumov, and
  Ronney]{FernandezGalisteo2018}
D.~Fernández-Galisteo, V.~N. Kurdyumov, and P.~D. Ronney.
\newblock Analysis of premixed flame propagation between two closely-spaced
  parallel plates.
\newblock \emph{Combust. Flame}, 190:\penalty0 133--145, 2018.
\newblock ISSN 0010-2180.

\bibitem[Creta et~al.(2020)Creta, Lapenna, Lamioni, Fogla, and
  Matalon]{Creta2020}
F.~Creta, P.~E. Lapenna, R.~Lamioni, N.~Fogla, and M.~Matalon.
\newblock Propagation of premixed flames in the presence of darrieus–landau
  and thermal diffusive instabilities.
\newblock \emph{Combust. Flame}, 216:\penalty0 256–270, June 2020.
\newblock ISSN 0010-2180.

\bibitem[Altantzis et~al.(2011)Altantzis, Frouzakis, Tomboulides, Kerkemeier,
  and Boulouchos]{Altantzis2011}
C.~Altantzis, C.~Frouzakis, A.~Tomboulides, S.~Kerkemeier, and K.~Boulouchos.
\newblock Detailed numerical simulations of intrinsically unstable
  two-dimensional planar lean premixed hydrogen/air flames.
\newblock \emph{Proc. Combust. Inst.}, 33:\penalty0 1261–1268, 2011.
\newblock ISSN 1540-7489.

\bibitem[Altantzis et~al.(2012)Altantzis, Frouzakis, Tomboulides, Matalon, and
  Boulouchos]{Altantzis2012}
C.~Altantzis, C.~E. Frouzakis, A.~G. Tomboulides, M.~Matalon, and
  K.~Boulouchos.
\newblock Hydrodynamic and thermodiffusive instability effects on the evolution
  of laminar planar lean premixed hydrogen flames.
\newblock \emph{J. Fluid Mech.}, 700:\penalty0 329–361, May 2012.
\newblock ISSN 1469-7645.

\bibitem[Frouzakis et~al.(2015)Frouzakis, Fogla, Tomboulides, Altantzis, and
  Matalon]{Frouzakis2015}
C.~E. Frouzakis, N.~Fogla, A.~G. Tomboulides, C.~Altantzis, and M.~Matalon.
\newblock Numerical study of unstable hydrogen/air flames: Shape and
  propagation speed.
\newblock \emph{Proc. Combust. Inst.}, 35:\penalty0 1087–1095, 2015.
\newblock ISSN 1540-7489.

\bibitem[Berger et~al.(2019)Berger, Kleinheinz, Attili, and Pitsch]{Berger2019}
L.~Berger, K.~Kleinheinz, A.~Attili, and H.~Pitsch.
\newblock Characteristic patterns of thermodiffusively unstable premixed lean
  hydrogen flames.
\newblock \emph{Proc. Combust. Inst.}, 37:\penalty0 1879–1886, 2019.
\newblock ISSN 1540-7489.

\bibitem[Attili et~al.(2021)Attili, Lamioni, Berger, Kleinheinz, Lapenna,
  Pitsch, and Creta]{Attili2021}
A.~Attili, R.~Lamioni, L.~Berger, K.~Kleinheinz, P.~E. Lapenna, H.~Pitsch, and
  F.~Creta.
\newblock The effect of pressure on the hydrodynamic stability limit of
  premixed flames.
\newblock \emph{Proc. Combust. Inst.}, 38:\penalty0 1973–1981, 2021.
\newblock ISSN 1540-7489.

\bibitem[Howarth and Aspden(2022)]{Howarth2022}
T.~Howarth and A.~Aspden.
\newblock An empirical characteristic scaling model for freely-propagating lean
  premixed hydrogen flames.
\newblock \emph{Combust. Flame}, 237:\penalty0 111805, March 2022.
\newblock ISSN 0010-2180.

\bibitem[Wen et~al.(2022{\natexlab{a}})Wen, Zirwes, Scholtissek, B\"{o}ttler,
  Zhang, Bockhorn, and Hasse]{Wen2022a}
X.~Wen, T.~Zirwes, A.~Scholtissek, H.~B\"{o}ttler, F.~Zhang, H.~Bockhorn, and
  C.~Hasse.
\newblock Flame structure analysis and composition space modeling of
  thermodiffusively unstable premixed hydrogen flames — {Part} i: Atmospheric
  pressure.
\newblock \emph{Combust. Flame}, 238:\penalty0 111815, April
  2022{\natexlab{a}}.
\newblock ISSN 0010-2180.

\bibitem[Wen et~al.(2022{\natexlab{b}})Wen, Zirwes, Scholtissek, B\"{o}ttler,
  Zhang, Bockhorn, and Hasse]{Wen2022b}
X.~Wen, T.~Zirwes, A.~Scholtissek, H.~B\"{o}ttler, F.~Zhang, H.~Bockhorn, and
  C.~Hasse.
\newblock Flame structure analysis and composition space modeling of
  thermodiffusively unstable premixed hydrogen flames — {Part} ii: Elevated
  pressure.
\newblock \emph{Combust. Flame}, 238:\penalty0 111808, April
  2022{\natexlab{b}}.
\newblock ISSN 0010-2180.

\bibitem[B\"{o}ttler et~al.(2024)B\"{o}ttler, Kaddar, Karpowski, Ferraro,
  Scholtissek, Nicolai, and Hasse]{Boettler2024}
H.~B\"{o}ttler, D.~Kaddar, T.~J.~P. Karpowski, F.~Ferraro, A.~Scholtissek,
  H.~Nicolai, and C.~Hasse.
\newblock Can flamelet manifolds capture the interactions of thermo-diffusive
  instabilities and turbulence in lean hydrogen flames?—an a-priori analysis.
\newblock \emph{Int. J. Hydrogen Energ.}, 56:\penalty0 1397–1407, February
  2024.
\newblock ISSN 0360-3199.

\bibitem[Schuh et~al.(2024)Schuh, Hasse, and Nicolai]{Schuh2024}
V.~Schuh, C.~Hasse, and H.~Nicolai.
\newblock An extension of the artificially thickened flame approach for
  premixed hydrogen flames with intrinsic instabilities.
\newblock \emph{Proc. Combust. Inst.}, 40:\penalty0 105673, 2024.
\newblock ISSN 1540-7489.

\bibitem[Pitsch(2024)]{Pitsch2024}
H.~Pitsch.
\newblock The transition to sustainable combustion: Hydrogen- and carbon-based
  future fuels and methods for dealing with their challenges.
\newblock \emph{Proc. Combust. Inst.}, 40:\penalty0 105638, 2024.
\newblock ISSN 1540-7489.

\bibitem[Altantzis et~al.(2015)Altantzis, Frouzakis, Tomboulides, and
  Boulouchos]{Altantzis2015}
C.~Altantzis, C.~E. Frouzakis, A.~G. Tomboulides, and K.~Boulouchos.
\newblock Direct numerical simulation of circular expanding premixed flames in
  a lean quiescent hydrogen-air mixture: Phenomenology and detailed flame front
  analysis.
\newblock \emph{Combust. Flame}, 162:\penalty0 331–344, February 2015.
\newblock ISSN 0010-2180.

\bibitem[Matalon et~al.(2003)Matalon, Cui, and Bechtold]{Matalon2003}
M.~Matalon, C.~Cui, and J.~K. Bechtold.
\newblock Hydrodynamic theory of premixed flames: effects of stoichiometry,
  variable transport coefficients and arbitrary reaction orders.
\newblock \emph{J. Fluid Mech.}, 487:\penalty0 179–210, June 2003.
\newblock ISSN 1469-7645.

\bibitem[Lapenna et~al.(2019)Lapenna, Lamioni, Troiani, and Creta]{Lapenna2019}
P.~E. Lapenna, R.~Lamioni, G.~Troiani, and F.~Creta.
\newblock Large scale effects in weakly turbulent premixed flames.
\newblock \emph{Proc. Combust. Inst.}, 37:\penalty0 1945–1952, 2019.
\newblock ISSN 1540-7489.

\bibitem[Lulic et~al.(2023)Lulic, Breicher, Scholtissek, Lapenna, Dreizler,
  Creta, Hasse, Geyer, and Ferraro]{Lulic2023}
H.~Lulic, A.~Breicher, A.~Scholtissek, P.~E. Lapenna, A.~Dreizler, F.~Creta,
  C.~Hasse, D.~Geyer, and F.~Ferraro.
\newblock On polyhedral structures of lean methane/hydrogen bunsen flames:
  Combined experimental and numerical analysis.
\newblock \emph{Proc. Combust. Inst.}, 39:\penalty0 1977–1986, 2023.
\newblock ISSN 1540-7489.

\bibitem[Welch et~al.(2024)Welch, Erhard, Shi, Dreizler, and
  B\"{o}hm]{Welch2024}
C.~Welch, J.~Erhard, H.~Shi, A.~Dreizler, and B.~B\"{o}hm.
\newblock An experimental investigation of lean hydrogen flame instabilities in
  spark-ignition engines.
\newblock \emph{Proc. Combust. Inst.}, 40:\penalty0 105391, 2024.
\newblock ISSN 1540-7489.

\bibitem[Dreizler and B\"{o}hm(2015)]{Dreizler2015}
A.~Dreizler and B.~B\"{o}hm.
\newblock Advanced laser diagnostics for an improved understanding of premixed
  flame-wall interactions.
\newblock \emph{Proc. Combust. Inst.}, 35:\penalty0 37–64, 2015.
\newblock ISSN 1540-7489.

\bibitem[Lai et~al.(2018)Lai, Klein, and Chakraborty]{Lai2018}
J.~Lai, M.~Klein, and N.~Chakraborty.
\newblock Direct numerical simulation of head-on quenching of statistically
  planar turbulent premixed methane-air flames using a detailed chemical
  mechanism.
\newblock \emph{Flow. Turbul. Combust.}, 101:\penalty0 1073–1091, April 2018.
\newblock ISSN 1573-1987.

\bibitem[Steinhausen et~al.(2023)Steinhausen, Zirwes, Ferraro, Scholtissek,
  Bockhorn, and Hasse]{Steinhausen2023}
M.~Steinhausen, T.~Zirwes, F.~Ferraro, A.~Scholtissek, H.~Bockhorn, and
  C.~Hasse.
\newblock Flame-vortex interaction during turbulent side-wall quenching and its
  implications for flamelet manifolds.
\newblock \emph{Proc. Combust. Inst.}, 39:\penalty0 2149–2158, 2023.
\newblock ISSN 1540-7489.

\bibitem[Fritz et~al.(2004)Fritz, Kröner, and Sattelmayer]{Fritz2004}
J.~Fritz, M.~Kröner, and T.~Sattelmayer.
\newblock Flashback in a swirl burner with cylindrical premixing zone.
\newblock \emph{J. Eng. Gas Turb. Power}, 126:\penalty0 276–283, April 2004.
\newblock ISSN 1528-8919.

\bibitem[Dabireau et~al.(2003)Dabireau, Cuenot, Vermorel, and
  Poinsot]{Dabireau2003}
F.~Dabireau, B.~Cuenot, O.~Vermorel, and T.~Poinsot.
\newblock Interaction of flames of {H$_{2}$} + {O$_{2}$} with inert walls.
\newblock \emph{Combust. Flame}, 135:\penalty0 123–133, October 2003.
\newblock ISSN 0010-2180.

\bibitem[Gruber et~al.(2010)Gruber, Sankaran, Hawkes, and Chen]{Gruber2010}
A.~Gruber, R.~Sankaran, E.~R. Hawkes, and J.~H. Chen.
\newblock Turbulent flame–wall interaction: a direct numerical simulation
  study.
\newblock \emph{J. Fluid Mech.}, 658:\penalty0 5–32, August 2010.
\newblock ISSN 1469-7645.

\bibitem[De~Nardi et~al.(2024)De~Nardi, Douasbin, Vermorel, and
  Poinsot]{DeNardi2024}
L.~De~Nardi, Q.~Douasbin, O.~Vermorel, and T.~Poinsot.
\newblock Infinitely fast heterogeneous catalysis model for premixed hydrogen
  flame-wall interaction.
\newblock \emph{Combust. Flame}, 261:\penalty0 113328, March 2024.
\newblock ISSN 0010-2180.

\bibitem[Mari et~al.(2016)Mari, Cuenot, Rocchi, Selle, and Duchaine]{Mari2016}
R.~Mari, B.~Cuenot, J.-P. Rocchi, L.~Selle, and F.~Duchaine.
\newblock Effect of pressure on hydrogen/oxygen coupled flame–wall
  interaction.
\newblock \emph{Combust. Flame}, 168:\penalty0 409–419, June 2016.
\newblock ISSN 0010-2180.

\bibitem[Zhao et~al.(2022)Zhao, Hernández~Pérez, Guo, Im, and Wang]{Zhao2022}
D.~Zhao, F.~E. Hernández~Pérez, C.~Guo, H.~G. Im, and L.~Wang.
\newblock Near wall effects on the premixed head-on hydrogen/air flame.
\newblock \emph{Combust. Flame}, 244:\penalty0 112267, October 2022.
\newblock ISSN 0010-2180.

\bibitem[Zhu et~al.(2024)Zhu, Zirwes, Zhang, Li, Zhang, and Pan]{Zhu2024}
J.~Zhu, T.~Zirwes, F.~Zhang, Z.~Li, Y.~Zhang, and J.~Pan.
\newblock Correlation of wall heat loss with quenching distance for premixed
  {H$_{2}$}/air flames during unsteady flame-wall interaction.
\newblock \emph{Chem. Eng. Science}, 283:\penalty0 119391, January 2024.
\newblock ISSN 0009-2509.

\bibitem[Chi et~al.(2024)Chi, Yu, Cuenot, Maas, and Thévenin]{Chi2024}
C.~Chi, C.~Yu, B.~Cuenot, U.~Maas, and D.~Thévenin.
\newblock Effect of differential diffusion on head-on quenching of premixed
  {NH$_{3}$}/{H$_{2}$}/air flames within turbulent boundary layers.
\newblock \emph{Proc. Combust. Inst.}, 40:\penalty0 105276, 2024.
\newblock ISSN 1540-7489.

\bibitem[Ojo et~al.(2024)Ojo, Padhiary, and Peterson]{Ojo2024}
A.~O. Ojo, A.~Padhiary, and B.~Peterson.
\newblock Spatiotemporal surface temperature measurements resolving flame-wall
  interactions of lean {H$_{2}$}-air and {CH$_{4}$}-air flames using phosphor
  thermometry.
\newblock \emph{Flow. Turbul. Combust.}, August 2024.
\newblock ISSN 1573-1987.

\bibitem[Weller et~al.(1998)Weller, Tabor, Jasak, and Fureby]{Weller1998}
H.~G. Weller, G.~Tabor, H.~Jasak, and C.~Fureby.
\newblock A tensorial approach to computational continuum mechanics using
  object-oriented techniques.
\newblock \emph{Comput. Phys.}, 12:\penalty0 620–631, November 1998.
\newblock ISSN 0894-1866.

\bibitem[Schneider et~al.(2024)Schneider, Steinhausen, Nicolai, and
  Hasse]{Schneider2024}
M.~Schneider, M.~Steinhausen, H.~Nicolai, and C.~Hasse.
\newblock Modeling of effusion cooling air-flame interaction using
  thermochemical manifolds.
\newblock \emph{Proc. Combust. Inst.}, 40:\penalty0 105453, 2024.
\newblock ISSN 1540-7489.

\bibitem[Li et~al.(2004)Li, Zhao, Kazakov, and Dryer]{Li2004}
J.~Li, Z.~Zhao, A.~Kazakov, and F.~L. Dryer.
\newblock An updated comprehensive kinetic model of hydrogen combustion.
\newblock \emph{Int. J. Chem. Kinet.}, 36:\penalty0 566–575, August 2004.
\newblock ISSN 1097-4601.

\bibitem[Kee et~al.(2017)Kee, Coltrin, Glarborg, and Zhu]{Kee2017}
R.~J. Kee, M.~E. Coltrin, P.~Glarborg, and H.~Zhu.
\newblock \emph{Chemically Reacting Flow: Theory, Modeling, and Simulation}.
\newblock Wiley, September 2017.
\newblock ISBN 9781119186304.

\bibitem[Goodwin et~al.(2023)Goodwin, Moffat, Schoegl, Speth, and
  Weber]{cantera}
D.~G. Goodwin, H.~K. Moffat, I.~Schoegl, R.~L. Speth, and B.~W. Weber.
\newblock Cantera: An object-oriented software toolkit for chemical kinetics,
  thermodynamics, and transport processes.
\newblock \url{https://www.cantera.org}, 2023.
\newblock Version 3.0.0.

\bibitem[Chapman and Cowling(1990)]{Chapman1990}
S.~Chapman and T.~Cowling.
\newblock \emph{The Mathematical Theory of Non-uniform Gases: An Account of the
  Kinetic Theory of Viscosity, Thermal Conduction and Diffusion in Gases}.
\newblock Cambridge Mathematical Library. Cambridge University Press, 1990.
\newblock ISBN 9780521408448.

\bibitem[Schlup and Blanquart(2017)]{Schlup2017}
J.~Schlup and G.~Blanquart.
\newblock Validation of a mixture-averaged thermal diffusion model for premixed
  lean hydrogen flames.
\newblock \emph{Combust. Theor. Model.}, 22:\penalty0 264–290, November 2017.
\newblock ISSN 1741-3559.

\bibitem[Hindmarsh et~al.(2005)Hindmarsh, Brown, Grant, Lee, Serban, Shumaker,
  and Woodward]{hindmarsh2005sundials}
A.~C. Hindmarsh, P.~N. Brown, K.~E. Grant, S.~L. Lee, R.~Serban, D.~E.
  Shumaker, and C.~S. Woodward.
\newblock {SUNDIALS}: Suite of nonlinear and differential/algebraic equation
  solvers.
\newblock \emph{ACM Transactions on Mathematical Software (TOMS)}, 31:\penalty0
  363--396, 2005.

\bibitem[Gardner et~al.(2022)Gardner, Reynolds, Woodward, and
  Balos]{gardner2022sundials}
D.~J. Gardner, D.~R. Reynolds, C.~S. Woodward, and C.~J. Balos.
\newblock Enabling new flexibility in the {SUNDIALS} suite of nonlinear and
  differential/algebraic equation solvers.
\newblock \emph{ACM Transactions on Mathematical Software (TOMS)}, 2022.

\bibitem[Niemeyer et~al.(2017)Niemeyer, Curtis, and Sung]{pyJac}
K.~E. Niemeyer, N.~J. Curtis, and C.-J. Sung.
\newblock pyjac: Analytical jacobian generator for chemical kinetics.
\newblock \emph{Compu. Phys. Commun.}, 215:\penalty0 188–203, June 2017.
\newblock ISSN 0010-4655.

\bibitem[Tekg\"{u}l et~al.(2021)Tekg\"{u}l, Peltonen, Kahila, Kaario, and
  Vuorinen]{Tekgl2021}
B.~Tekg\"{u}l, P.~Peltonen, H.~Kahila, O.~Kaario, and V.~Vuorinen.
\newblock Dlbfoam: An open-source dynamic load balancing model for fast
  reacting flow simulations in openfoam.
\newblock \emph{Compu. Phys. Commun.}, 267:\penalty0 108073, October 2021.
\newblock ISSN 0010-4655.

\bibitem[Poinsot et~al.(1993)Poinsot, Haworth, and Bruneaux]{Poinsot1993}
T.~Poinsot, D.~Haworth, and G.~Bruneaux.
\newblock Direct simulation and modeling of flame-wall interaction for premixed
  turbulent combustion.
\newblock \emph{Combust. Flame}, 95:\penalty0 118–132, October 1993.
\newblock ISSN 0010-2180.

\end{thebibliography}

\end{document}